\documentclass[a4paper,12pt]{article}
\usepackage{amsfonts}
\usepackage{smart}
\usepackage{float}
\usepackage{latexsym}
\link{equation}{section}\toheight1
\newcommand{\clth}{\setcounter{thm}{0}}
\newcommand{\sectionnew}[1]{\section{#1}\clth}
\newtheorem{thm}{Theorem}[section]\toheight1
\newcommand{\beq}{\begin{equation}}
\newcommand{\eeq}{\end{equation}}
\newcommand{\beqa}{\begin{eqnarray}}
\newcommand{\eeqa}{\end{eqnarray}}
\newcommand{\nn}{\nonumber \\}
\newtheorem{prop}[thm]{Proposition}
\newcommand {\np}[1]{ {\mathrm{:}}{#1}{\mathrm{:}} } %norm.pr
\newcommand {\sg}[1]{ {\mathrm{sgn}}({#1}) }
\newcommand {\I}[1]{ {\mathrm{Im}}\, {#1} }
\def \a {\underline{\alpha}}
\def \l {\lambda}
\def \L {\Lambda}
\def \s {\sigma}
\def \r {\rho}
\def \e {{\bf{e}}}
\def \Q {{\bf{Q}}}
\def \q {{\bf q}}
\def \u {{\bf{u}}}
\def \v {{\bf{v}}}
\def \ex {\mathrm{e}}
\def \ep {\varepsilon}
\def \h {\frac{1}{2}}
\def \z {\zeta}

\def \W {W_{1+\infty}\ }
\def \A {{\mathcal A} }
\def \d {\delta}
\def \D {\Delta}

\def \la {\langle}
\def \ra {\rangle}
\def \V {{\mathcal V}}

\def \T {{\mathcal T}}

\def \H {{\mathcal H}}
\def \Lc {{\mathcal L}}
\def \Rset {{\mathbb R}}

\def \Z {{\mathbb Z}}
\def \N {{\mathbb N}}
\def \G   {\Gamma}
\def \dd {\mathrm{d}}
\def \mod {\, \mathrm{mod}\, }
\def\U1{{\widehat{u(1)}}}

\begin{document}
\vspace {-1.cm}
\title{A Unified Conformal Field Theory Description
of Paired Quantum Hall States}
\author{ A. Cappelli
\thanks{E-mail: andrea.cappelli@fi.infn.it}
\\
\normalsize\textit{ I.N.F.N. and Dipartimento di Fisica,}
\\
\normalsize\textit{ Largo E. Fermi 2, I-50125 Firenze, Italy}
\\
\and L. S. Georgiev
\thanks{E-mail: lgeorg@inrne.bas.bg}
\qquad I. T. Todorov
\thanks{E-mail: todorov@inrne.bas.bg, itodorov@esi.ac.at}
\\
\normalsize\textit{E. Schr\"{o}dinger Inst. for Mathematical Physics,} \\
\normalsize\textit{ Boltzmanngasse 9, A-1090 Wien, Austria} \\
\normalsize\textit{and }\\
\normalsize\textit{ Institute for Nuclear
Research and Nuclear Energy,}\\
\normalsize\textit{ Tsarigradsko Chaussee 72,
BG-1784 Sofia, Bulgaria }
}
\date{}
\maketitle
\vspace {-.7cm}

\begin{abstract}
The wave functions of the Haldane-Rezayi paired Hall state
have been previously described by a non-unitary conformal field
theory with central charge $c=-2$.
Moreover, a relation with the $c=1$ unitary Weyl fermion has been suggested.
We construct the complete unitary theory and show that it
consistently describes the edge excitations of the Haldane-Rezayi state.
Actually, we show that the unitary ($c=1$) and non-unitary ($c=-2$) theories 
are related by a local map between the two sets of fields
and by a suitable change of conjugation.
The unitary theory of the Haldane-Rezayi state is found to be
the same as that of the 331 paired Hall state.
Furthermore, the analysis of modular invariant partition functions
shows that no alternative unitary descriptions are possible
for the Haldane-Rezayi state within the class of rational
conformal field theories with abelian current algebra.
Finally, the known $c=3/2$ conformal theory of the Pfaffian state
is also obtained from the 331 theory by a reduction of degrees of 
freedom which can be physically realized in the double-layer Hall systems.
\end{abstract}

\hfill Preprint ESI 621 (1998) and DFF 329/10/98, hep-th/9810105
\newpage
\tableofcontents
\newpage

%-1----------------------------

\section{Introduction}

The so-called  331 \cite{halp}, Pfaffian \cite{more} and
Haldane-Rezayi (HR) \cite{hr} $\nu=1/2$  $(5/2)$ quantum Hall (QH) states
\cite{prange} have been analysed extensively in the recent
literature. They are called {\it paired} Hall states \cite{gww} because they
contain two kinds of electrons, carrying spin or layer index,
which first bind in pairs and then form incompressible fluids
\cite{laugh}.

One would like to identify the Conformal Field Theories (CFT) \cite{cft}
corresponding to these states, which describe their low-energy  edge
excitations \cite{wen}. This requires some guesswork and ingenuity
for reconstructing the complete Hilbert space from the knowledge
of the ground-state wave function and possibly some quasi-particle states.
There are well-established procedures which have been used for the
spin-polarized single-layer Hall states \cite{wen,prange},
but they do not seem good enough for the paired states.
In particular, the CFT proposed
for the HR state is puzzling for the lack of unitarity
\cite{ww,mire,gfn}, or locality \cite{lw,lud}.

In this paper, we present a unified description of the paired Hall
states which uses the same conformal fields (or a subset of them) in all
three cases. We show that an unitary description of the
HR state is possible and that this is given by the same CFT as that of
the 331 state; furthermore, we interpret  the Pfaffian as a projection
of the 331 state, which can be obtained
in the limit of low potential barrier between the two layers
\cite{gww}.

This common CFT description is rather useful for
the physical interpretation; moreover, it allows the discussion of the
$\W$ symmetry of the paired states.
This symmetry characterizes the incompressible Hall fluids \cite{ctz1},
and is a definitive building criterion for the CFTs of the hierarchical
single-layer states \cite{ctz2}.
We show that the $\W$ symmetry also characterizes the 331 and HR
double-layer states, and that it is broken at the quantum
level in the Pfaffian state.

In view of the controversial literature on this subject, it is
important to state the hypotheses made in
this work: we consider rational conformal field theories (RCFT),
whose completeness can be checked  by computing  their modular
invariant partition functions \cite{cz}; we require the unitarity
of the theories, because they describe physical excitations propagating
at the edge. Moreover, we consider, whenever possible,
theories with a (multi-component) abelian current algebra, which 
possess the $\W$ symmetry and can be extended to RCFTs \cite{kt}
(henceforth called {\it lattice} RCFTs).

Therefore, in this paper we specifically prove that there is only
one $c=2$ unitary lattice RCFT suitable
for the HR state -- that of the 331 state.
In particular, the HR ground state appears as an excited state in the 331 CFT.
This result is at variance with the common opinion 
that these Hall states identify two independent universality classes,
with different numerical energy spectrum \cite{hr,rere} 
and topological order \cite{hr,wkk}.
Our result may imply that this is not completely correct, or,
alternatively, that the HR state is not described by a lattice
RCFT; in either case, the unitarity problem is cleared up.
Finally, the Pfaffian state is consistently described in terms of the 
same 331 conformal fields.

%-1.1---------------------------

\subsection{Review of the Haldane-Rezayi State}

Here we review the basic characteristics of the model
and its available theoretical
treatments. The wave function $\Psi$ of a paired QH system of
$2N$ electrons is written as a product of the usual
Gaussian factor and an analytic function of the electron
coordinates $z_i$ and $w_i$ of the first and the second
layer, respectively (alternatively, of the up and down spin components):
\beq \label{1.1}
\Psi(z_i, w_i;\bar{z}_i,\bar{w}_i)=
\Phi(z_1,w_1, \ldots z_N,w_N)
\exp{\left(-\frac{1}{4}\sum_i (|z_i|^2 + |w_i|^2)\right)}.
\eeq
In the framework developed in \cite{more}\cite{ww} (see also
earlier work cited there),  the analytic factor $\Phi$ is interpreted
as the CFT correlation function: \\
$\langle  \Phi_N | \phi^1(z_1) \cdots \phi^1(z_N)
\phi^2(w_1) \cdots  \phi^2(w_N)|0\rangle $, where
$\phi^i$ is a chiral conformal field of
effective charge $\q^i \;\;(i=1,2)$ representing the
electrons of layer $i$, and $\la \Phi_N |$ is the out state 
carrying a compensating charge $N(\q^1+\q^2)$.
The holomorphic wave function of the HR state 
\cite{hr} is written as the product,
\beq \label{1.3}
\Phi_{HR}(z_i; w_i)= \Phi_m(z_i; w_i)\Phi_{ds}(z_i; w_i) \ ,
\eeq
of a Laughlin type wave-function \cite{laugh},
\beq \label{1.4}
\Phi_m(z_i; w_i)=\prod_{i < j}(z_{ij}w_{ij})^m \prod_{i,j}
(z_i-w_j)^m ,\;\;\; m=2,4, \ldots, \; z_{ij}=z_i-z_j\ ,
\eeq
and a neutral {\em d-wave spin-singlet} part, $\Phi_{ds}$, that is
skew-symmetric in $(z_1,\ldots , z_N)$ and in
$(w_1,\ldots , w_N)$, separately:
\beq\label{1.5}
\Phi_{ds}(z_i;w_i)= (-1)^{\frac{N(N-1)}{2}}
{\mathrm {det}} \left( \frac{1}{(z_i -w_j)^2} \right).
\eeq
This expression can be viewed, following Ref.\cite{ww}, as the 
$2N$-point vacuum expectation value of a pair of Fermi fields
$\psi_\pm(z)$ ($\psi_{\pm\h}(z)$, the subscript of $\psi$ referring,
alternatively to spin projection or layer):
\beq\label{1.6}
\Phi_{ds} (z_i;w_i) =
\la 0 \vert\psi_+(z_1) \cdots \psi_+(z_N)\ \psi_-(w_1)\cdots\psi_-(w_N)
\vert 0\ra \ .
\eeq
Equation (\ref{1.5}) would then
follow (in a local field theory with energy bounded from below):
(i) from the {\it ``quasi-free"} anti-commutation relations
\cite{ww,mire}:
\beq\label{1.7}
[\psi_\r(z),\psi_\s(w)]_+=-\ep_{\r\s}\d'(z-w)\ , \quad
\ep_{\r\s}=-  \ep_{\r\s}\ , \quad \ep_{+-}=1\ ,
\eeq
where $\d(z-w)$ is the Dirac delta function for holomorphic test
functions; and (ii) from the knowledge of the 2-point functions
(which restricts the choice of vacuum).
We adopt (\ref{1.7}) in what follows as a phenomenological input.

The knowledge of the $2N$-point function (\ref{1.6})
allows to determine the operator content of (the vacuum sector of)
the CFT generated by the pair $\psi_\pm(z)$.
To do that we write the determinant in Eq. (\ref{1.5}) in the form
\beqa
\det \left(\frac{1}{(z_i-w_j)^2} \right) &=&
\det \left(\frac{1}{z_i-w_j} \right)
\mathrm{perm} \left(\frac{1}{z_i-w_j} \right), \nn
\det \left(\frac{1}{z_i-w_j} \right) &=& \left(-1\right)^{N(N-1)/2}
\frac{\prod\limits_{i<j} z_{ij}w_{ij} }{\prod\limits_{i,j} (z_i-w_j)};
\nonumber
\eeqa
here the {\em permanent} is the symmetrized product of $(z_i-w_j)^{-1}$
which has a non-zero limit
\[ \mathrm{perm} \left(\frac{1}{z_i-w_j} \right)=
\sum_{\s\in S_N}\prod\limits_{i=1}^N \frac{1}{z_i-w_{\s(i)}}
\quad\rightarrow\quad \frac{N!}{(z-w)^N}  \ ,\]
for $ z_i\rightarrow z$, $w_j\rightarrow w$, $ i,j=1,\dots, N$.
These properties of the 
$2N$-point function (\ref{1.5}) imply that products of
$\psi_\pm(z)$ give rise to a sequence of
composite fields $V_{\pm s}$ ($V_{\pm \h}(z)=\psi_\pm(z)$)
of dimension $\D(2s)$ determined inductively by the
operator-product expansion (OPE):
\beq\label{1.8'}
 \psi_\pm(z_1)V_{\pm s}(z_2)\sim z_{12}^{2s}  \ V_{\pm s\pm\h}(z_2)\ ,
\qquad (2s=0,1,2,\dots)\ ,
\eeq
implying,
\beq\label{1.9'}
\D(\pm 2s\pm 1)-\D(\pm 2s)-\D(\pm 1)=2s \quad \Rightarrow \quad
\D(\pm 2s)=s\left(2s +2\D(\pm 1)-1\right).
\eeq
Eq. (\ref{1.8'})  allows to express $V_{\pm s}$
as (normal) products of $\psi_{\pm}$  and their derivatives:
\beq\label{1.11'}
V_{\pm s}(z)=\prod\limits_{i=1}^{2s} \np{
\frac{1}{(2s-i)!}\partial^{2s-i}\psi_\pm(z)} \ .
\eeq
The values $\D(\pm 1)$ depend on the choice of the stress energy
tensor; however, according to (\ref{1.5}) their sum is fixed:
\beq\label{1.12'}
\D(1)+\D(-1)=2 \ .
\eeq

In Ref. \cite{ww}, it was further assumed that $\psi_\pm$ are
{\it primary conformal fields}. This implies that these Fermi fields
have conformal dimension one in violation of the spin-statistics
relation for a unitary CFT.
Indeed, these fields can be found \cite{ww,mire,gfn}
in the non-unitary extension of the Virasoro
minimal conformal  models \cite{cft} with central charge
$c_p=1-6/(p(p+1))$, for the value $p=1$, -i.e., $c=-2$.
The non-unitary stress tensor $\T(z)=\np{\psi_- \psi_+}$ is then invariant
under the SU(2)-spin group and so are the
OPEs of any number of $\psi_\pm$ factors (see \cite{ww} as
well as Section 3 and Appendix B below). It follows that for each
$2s = 0,1,2,\dots$, the composite operators $V_{\pm s}$
(\ref{1.11'}) are the lowest and highest spin projection components
of a $(2s+1)$-dimensional multiplet of fields $\phi_{sm}(z)$,
$m=-s,-s+1,\ldots , s$, of the same dimension (\ref{1.9'})
(for $\D(1)=1$):
\beq\label{1.13'}
\D(\pm 2s)=s(2s+1) \qquad (2s=0,1,2,\ldots).
\eeq

This $c=-2$ CFT is known in the literature as the $\xi$-$\eta$
``ghost system" \cite{fms}, which is defined by the pair of canonical
Fermi fields,
\beqa\label{1.8a}
{}[\xi(z),\eta(w)]_+ &=& \d(z-w)\ , \nn
{}[\xi(z),\xi(w)]_+  &=&  0 \ = \ [\eta(z),\eta(w)]_+ \ ,\nn
\la 0\vert\eta(z)\xi(w)\vert 0\ra&=& \frac{1}{z-w} \ ,
\eeqa
of conformal dimensions $\D_\xi=0,\; \D_\eta=1$. 
Their normal product defines the current $j$ :
\beqa\label{1.8a'}
\eta(z)\xi(w) &=& \frac{1}{z-w}   +\np{\eta(z)\xi(w)}\ ,\\
j(z) &=& \np{\xi(z)\eta(z)}=\sum_{n\in\Z} j_n z^{-n-1}\ .
\label{1.8b}
\eeqa

We can set:
\beq\label{1.9}
\psi_-(z)=\xi'(z)\ ,\qquad  \psi_+(z)=\eta(z) \ .
\eeq
Note that the zero mode $j_0$ acts as (minus twice)
the spin projection operator $S_3$, which counts the difference
between spin up and spin down components  $\psi_\pm$:
$[\psi_\pm(z), j_0]\equiv [\psi_\pm(z),2 S_3]=\pm \psi_\pm(z) $, with
$\vec{S}=(S_1,S_2,S_3)$ standing for the spin operator.
Moreover, the stress tensor
$\T(z)=\np{\xi'(z)\eta(z)}$  assigns  dimension
one to $\psi_\pm$, and the central charge is $c=-2$.

According to the Kac determinant formula, the conformal dimensions
of the $c=-2$ primary fields are given by \cite{cft}:
\beq\label{1.10}
\D_l= {l^2 -1 \over 8} \ge - \frac{1}{8} \ , \qquad l=0,1,2,\dots
\eeq
In particular, for odd $l$, $l=4s+1$  $(2s=0,1,2,\dots)$,
we recover the spin multiplets of integer dimensions given by (\ref{1.13'}).
The state of lowest dimension in this theory is the {\em disorder state}
\cite{ww} with $\D_0=-1/8$, which is a spin singlet;
it gives rise to a $\Z_2$ twisted sector in
which the field  $\psi_\r$ creates double-valued quasi-particles
with dimensions $\D_{2n}= (n^2/2) - 1/8$.
In fact, it follows from (\ref{1.10}) that:
\beq\label{1.12}
\psi_\r(z) |\D_0\ra\sim z^{-\h} |\D_2 \ra \ ,\qquad
\left(\D_2-\D_0=\h\right)\ .
\eeq
The appearance of negative norm states, like $\T(0)|0\ra$,
 and a negative conformal
dimension ($\D_0=-\frac{1}{8}$) is certainly untenable from a physical
point of view.
This non-unitarity problem has been recognized and various solutions
have been proposed in Refs. \cite{mire,gfn,lw,lud}.

In this paper, we choose to relax the assumption that the fields
$\psi_\r$ are {\it primary}, while keeping, at the same time, the property
(\ref{1.12}) for the quasi-particle excitations.
We obtain a unitary theory of the HR model which is
based on the correspondence between the Hilbert space and the fields of
the $c=-2$ CFT and those of the $c=1$ Weyl fermion theory
(see also Refs.\cite{gfn,lw,lud}); we keep the
expression (\ref{1.3}-\ref{1.5}) of the HR wave function
while preserving both {\em conformal invariance } and {\em modular invariance}
\cite{cz,gan}.
Moreover,  we can define a {\em hermitean} stress tensor which,
however, is not SU(2) invariant.

%-1.2-------------------------------------

\subsection{Outline of the Paper}

In Section 2, we first introduce a $U(1)\times U(1)$
current algebra CFT, whose orthogonal lattice
contains a Weyl fermion field and a Laughlin boson.
We use it to describe the 331 CFT as a $\Z_2$ orbifold (in the sense
of \cite{kt}), whose ($\Z_2$-even) states possess
charge and fermion number coupled by the {\it parity (``projection'') rule}
defined in Ref.\cite{mire}.
The modular invariant partition function for the 331 RCFT is also obtained
in agreement with this rule.

In Section 3, we present the $c=-2$ to $c=1$ correspondence
for the HR theory: we first discuss the improvement of the $c=-2$
stress tensor which leads to a unitary theory and then map the fields
and the characters of the current-algebra representations.
Furthermore, we find that the $c=-2$ partition function of the HR model
proposed in Ref. \cite{mire} is mapped into the 331 one;
this shows that the CFT descriptions of the two models coincide,
once unitarity is enforced; moreover, we find that the 
parity rule is the same in the two theories.

In Section 4, we classify all possible $U(1)\times U(1)$ lattice
current algebras, which can be made with the excitations of the 331
and HR theories (assuming the standard charge-statistics relation for the
observable electron-like excitations).
We find that there is a unique modular invariant partition function:
thus, there is only one possible unitary lattice RCFT which can
describe the edge excitations of the HR state, which is the same as that of
the 331 state.

In Section 5, we show that the Pfaffian state can be obtained from 
the 331 model by a {\it generalized gauge reduction}:
namely, its $c=3/2$ CFT of a Majorana fermion and a Laughlin boson
\cite{more} is reproduced by projecting out the imaginary part
of the Weyl fermion of the 331 CFT.
The corresponding reduction at the level of partition functions
shows that the Pfaffian theory inherits the parity rule of the
two other paired Hall states.
This gauge reduction breaks the $\W$ symmetry present in the
331 theory and gives  rise to  non-Abelian statistics
for the quasi-particles.
In Section 5.2, a similar reduction allows to relate the
maximally-symmetric $SU(2)\times SU(2)$ $c=3$ lattice RCFT
of Ref.\cite{fst} with the 331 CFT.

In the Conclusions (Section 6), we discuss some problems left open by our
analysis, most notably the possible ways to distinguish between the
331 and the HR states.
The Appendices contain more technical discussions:
Appendix A sums up some basic facts about charge lattices and orbifolds
of finite cyclic groups needed in the text.
In Appendix B we prove that the $SU(2)$ invariant OPE of
$\psi_\rho (z) \psi_\s (w)$ is independent of the choice of the stress
tensor and of the dimensions of these fields.
Finally, Appendix C provides a complete list of  modular
invariants of the orthogonal lattice algebra (Section 2) underlying both
the 331 and the HR states. This is used in Section 4 to show
that there is a unique lattice RCFT for the 331 and HR states.

%-2------------------------------

\sectionnew{The 331 Model as a $\Z_2$ Orbifold of an
Orthogonal Lattice Algebra}

\subsection{The $(m+1)(m+1)(m-1)$ Holomorphic Wave Function and the Associated
Charge Lattice }

\begin{figure}[H]
\begin{center}
\caption{\label{fig.1}The original lattice $L$ (encircled dots)
         as a sub-lattice of the orthogonal  $\G_{m,1}$ (dots)}
%%%
\setlength{\unitlength}{0.240900pt}
\ifx\plotpoint\undefined\newsavebox{\plotpoint}\fi
\begin{picture}(1049,900)(0,0)
\font\gnuplot=cmr10 at 10pt
\gnuplot
\sbox{\plotpoint}{\rule[-0.200pt]{0.400pt}{0.400pt}}%
\put(176.0,473.0){\rule[-0.200pt]{194.888pt}{0.400pt}}
\put(581.0,68.0){\rule[-0.200pt]{0.400pt}{194.888pt}}
\put(176.0,149.0){\rule[-0.200pt]{4.818pt}{0.400pt}}
\put(154,149){\makebox(0,0)[r]{-4}}
\put(965.0,149.0){\rule[-0.200pt]{4.818pt}{0.400pt}}
\put(176.0,311.0){\rule[-0.200pt]{4.818pt}{0.400pt}}
\put(154,311){\makebox(0,0)[r]{-2}}
\put(965.0,311.0){\rule[-0.200pt]{4.818pt}{0.400pt}}
\put(176.0,473.0){\rule[-0.200pt]{4.818pt}{0.400pt}}
\put(154,473){\makebox(0,0)[r]{0}}
\put(965.0,473.0){\rule[-0.200pt]{4.818pt}{0.400pt}}
\put(176.0,634.0){\rule[-0.200pt]{4.818pt}{0.400pt}}
\put(154,634){\makebox(0,0)[r]{2}}
\put(965.0,634.0){\rule[-0.200pt]{4.818pt}{0.400pt}}
\put(176.0,796.0){\rule[-0.200pt]{4.818pt}{0.400pt}}
\put(154,796){\makebox(0,0)[r]{4}}
\put(965.0,796.0){\rule[-0.200pt]{4.818pt}{0.400pt}}
\put(257.0,68.0){\rule[-0.200pt]{0.400pt}{4.818pt}}
\put(257,23){\makebox(0,0){-4}}
\put(257.0,857.0){\rule[-0.200pt]{0.400pt}{4.818pt}}
\put(419.0,68.0){\rule[-0.200pt]{0.400pt}{4.818pt}}
\put(419,23){\makebox(0,0){-2}}
\put(419.0,857.0){\rule[-0.200pt]{0.400pt}{4.818pt}}
\put(581.0,68.0){\rule[-0.200pt]{0.400pt}{4.818pt}}
\put(581,23){\makebox(0,0){0}}
\put(581.0,857.0){\rule[-0.200pt]{0.400pt}{4.818pt}}
\put(742.0,68.0){\rule[-0.200pt]{0.400pt}{4.818pt}}
\put(742,23){\makebox(0,0){2}}
\put(742.0,857.0){\rule[-0.200pt]{0.400pt}{4.818pt}}
\put(904.0,68.0){\rule[-0.200pt]{0.400pt}{4.818pt}}
\put(904,23){\makebox(0,0){4}}
\put(904.0,857.0){\rule[-0.200pt]{0.400pt}{4.818pt}}
\put(176.0,68.0){\rule[-0.200pt]{194.888pt}{0.400pt}}
\put(985.0,68.0){\rule[-0.200pt]{0.400pt}{194.888pt}}
\put(176.0,877.0){\rule[-0.200pt]{194.888pt}{0.400pt}}
\put(686,501){\makebox(0,0)[l]{$\mathrm{e}^1$}}
\put(593,561){\makebox(0,0)[l]{$\mathrm{e}^2$}}
\put(710,574){\makebox(0,0)[l]{$\mathrm{q}^1$}}
\put(706,375){\makebox(0,0)[l]{$\mathrm{q}^2$}}
\put(176.0,68.0){\rule[-0.200pt]{0.400pt}{194.888pt}}
\put(581,473){\vector(1,0){114}}
\put(581,473){\vector(0,1){80}}
\multiput(581.00,473.58)(0.713,0.499){157}{\rule{0.670pt}{0.120pt}}
\multiput(581.00,472.17)(112.609,80.000){2}{\rule{0.335pt}{0.400pt}}
\put(695,553){\vector(3,2){0}}
\multiput(581.00,471.92)(0.704,-0.499){159}{\rule{0.663pt}{0.120pt}}
\multiput(581.00,472.17)(112.624,-81.000){2}{\rule{0.331pt}{0.400pt}}
\put(695,392){\vector(4,-3){0}}
\put(237,68){\circle{24}}
\put(237,230){\circle{24}}
\put(352,149){\circle{24}}
\put(466,68){\circle{24}}
\put(237,392){\circle{24}}
\put(352,311){\circle{24}}
\put(466,230){\circle{24}}
\put(581,149){\circle{24}}
\put(695,68){\circle{24}}
\put(237,553){\circle{24}}
\put(352,473){\circle{24}}
\put(466,392){\circle{24}}
\put(581,311){\circle{24}}
\put(695,230){\circle{24}}
\put(809,149){\circle{24}}
\put(924,68){\circle{24}}
\put(237,715){\circle{24}}
\put(352,634){\circle{24}}
\put(466,553){\circle{24}}
\put(581,473){\circle{24}}
\put(695,392){\circle{24}}
\put(809,311){\circle{24}}
\put(924,230){\circle{24}}
\put(237,877){\circle{24}}
\put(352,796){\circle{24}}
\put(466,715){\circle{24}}
\put(581,634){\circle{24}}
\put(695,553){\circle{24}}
\put(809,473){\circle{24}}
\put(924,392){\circle{24}}
\put(466,877){\circle{24}}
\put(581,796){\circle{24}}
\put(695,715){\circle{24}}
\put(809,634){\circle{24}}
\put(924,553){\circle{24}}
\put(695,877){\circle{24}}
\put(809,796){\circle{24}}
\put(924,715){\circle{24}}
\put(924,877){\circle{24}}
\put(237,68){\circle*{12}}
\put(237,149){\circle*{12}}
\put(237,230){\circle*{12}}
\put(237,311){\circle*{12}}
\put(237,392){\circle*{12}}
\put(237,473){\circle*{12}}
\put(237,553){\circle*{12}}
\put(237,634){\circle*{12}}
\put(237,715){\circle*{12}}
\put(237,796){\circle*{12}}
\put(237,877){\circle*{12}}
\put(352,68){\circle*{12}}
\put(352,149){\circle*{12}}
\put(352,230){\circle*{12}}
\put(352,311){\circle*{12}}
\put(352,392){\circle*{12}}
\put(352,473){\circle*{12}}
\put(352,553){\circle*{12}}
\put(352,634){\circle*{12}}
\put(352,715){\circle*{12}}
\put(352,796){\circle*{12}}
\put(352,877){\circle*{12}}
\put(466,68){\circle*{12}}
\put(466,149){\circle*{12}}
\put(466,230){\circle*{12}}
\put(466,311){\circle*{12}}
\put(466,392){\circle*{12}}
\put(466,473){\circle*{12}}
\put(466,553){\circle*{12}}
\put(466,634){\circle*{12}}
\put(466,715){\circle*{12}}
\put(466,796){\circle*{12}}
\put(466,877){\circle*{12}}
\put(581,68){\circle*{12}}
\put(581,149){\circle*{12}}
\put(581,230){\circle*{12}}
\put(581,311){\circle*{12}}
\put(581,392){\circle*{12}}
\put(581,473){\circle*{12}}
\put(581,553){\circle*{12}}
\put(581,634){\circle*{12}}
\put(581,715){\circle*{12}}
\put(581,796){\circle*{12}}
\put(581,877){\circle*{12}}
\put(695,68){\circle*{12}}
\put(695,149){\circle*{12}}
\put(695,230){\circle*{12}}
\put(695,311){\circle*{12}}
\put(695,392){\circle*{12}}
\put(695,473){\circle*{12}}
\put(695,553){\circle*{12}}
\put(695,634){\circle*{12}}
\put(695,715){\circle*{12}}
\put(695,796){\circle*{12}}
\put(695,877){\circle*{12}}
\put(809,68){\circle*{12}}
\put(809,149){\circle*{12}}
\put(809,230){\circle*{12}}
\put(809,311){\circle*{12}}
\put(809,392){\circle*{12}}
\put(809,473){\circle*{12}}
\put(809,553){\circle*{12}}
\put(809,634){\circle*{12}}
\put(809,715){\circle*{12}}
\put(809,796){\circle*{12}}
\put(809,877){\circle*{12}}
\put(924,68){\circle*{12}}
\put(924,149){\circle*{12}}
\put(924,230){\circle*{12}}
\put(924,311){\circle*{12}}
\put(924,392){\circle*{12}}
\put(924,473){\circle*{12}}
\put(924,553){\circle*{12}}
\put(924,634){\circle*{12}}
\put(924,715){\circle*{12}}
\put(924,796){\circle*{12}}
\put(924,877){\circle*{12}}
\end{picture}
%%%
\end{center}
\end{figure}
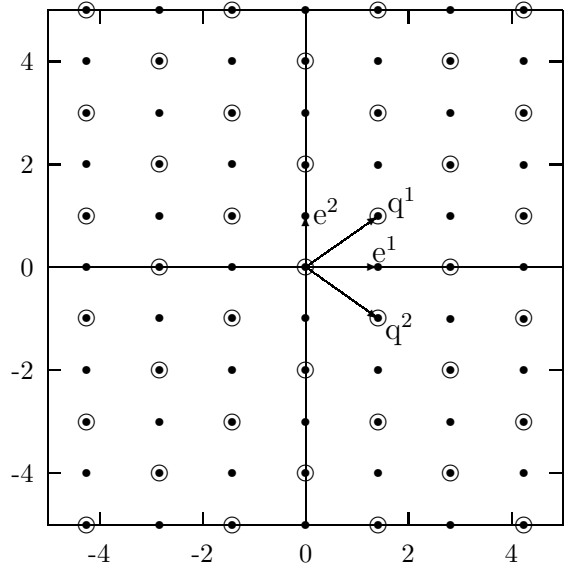

We shall be dealing in this section with a natural
generalization of the 331  model corresponding to the  filling fraction
 $\nu=1/m$,  $m$ even, and to  the holomorphic wave function:
\beqa\label{1.13}
\Phi_{(m+1)(m+1)(m-1)}\left(z_i;w_i\right)&=&
\prod_{1\leq i<j\leq N}
\left(z_{ij}w_{ij}\right)^{m+1}  \prod_{i,j} \left(z_i-w_j\right)^{m-1} \nn
&=&\left(-1\right)^{N(N-1)/2} \Phi_m\left( z_i;w_i\right)
\det\left (\frac{1}{z_i-w_j}\right) ,
\eeqa
where $\Phi_m$ is the $U(1)$ factor (\ref{1.4}).
This ground state wave function is reproduced in a $c=2$ RCFT whose
chiral algebra $\A(L)$ is an extension of the $\U1\times\U1$
current algebra by two pairs of oppositely charged fields of charge vectors
$\q^1$ and $\q^2$ spanning a two-dimensional lattice $L$.
The Gram matrix of $L$ is
\beq\label{3.7}
G_{L}=\left((\q^i|\q^j) \right)=
\left(\matrix{
m+1 & m-1 \cr
m-1 & m+1 \cr}\right)\ ,\qquad \left( m=2,4,\dots \right)\ .
\eeq
The resulting RCFT, called the $(m+1)(m+1)(m-1)$ model,
can be constructed from the following observation:
we can embed the lattice $L$ in a finer, orthogonal
one,  such that the corresponding conformal theory is the
direct product of a Weyl fermion and a Laughlin anyon with
$\nu=1/m$ (Fig. \ref{fig.1}).
This basis will give the natural description for the quasi-particle
excitations of all the paired Hall states, which
will only differ in the treatment of the neutral fermionic factor.

\begin{prop}
(a) The lattice $L$ is a sub-lattice of index two of the (integral)
orthogonal lattice $\G_{m,1} = \Z \e^1 \oplus \Z \e^2$
with Gram matrix:
\beq\label{3.1}
G_{m,1}=\left((\e^i|\e^j) \right)=
\left(\matrix{
m & 0 \cr
0 & 1 \cr}\right) \ .
\eeq
(b) We associate with the lattice $L$ a $\U1 \otimes\U1$
chiral current algebra
$\A_{m,1}$, by letting the vectors $\pm \e^1$ correspond to
a pair of oppositely charged Bose fields $E^g (z)$ ,
$g=\pm\sqrt{m}$, of  dimension $m/2 \in \N$, and $\pm \e^2$ to a pair of
conjugate Weyl fermions $(\psi(z), \psi^*(z))$ of dimension $1/2$.
Then, the algebra $\A_{m,1}$ admits an involutive inner
automorphism $\alpha$ defined on the generating fields by:
\beq\label{action}
\alpha\left[ E^g (z) \right] = - E^g (z) \ , \qquad
\alpha \left[ \psi^{(*)} (z) \right] = - \psi^{(*)}(z) \ ,\qquad
g=\pm\sqrt{m} \ ,
\eeq
while $\A(L)$ appears as the sub-algebra of $\Z_2$
 invariant elements of $\A_{m,1}$ (the fixed points of $\alpha$).
It is generated by the products of commuting fields:
\beq\label{3.5}
\psi^1(z)=\psi^*(z)E^{-\sqrt m}(z)\ , \qquad
\psi^2(z)=\psi(z)E^{-\sqrt m}(z) \ .
\eeq
\end{prop}

{\bf Proof.} (a) The lattice  $L= \Z \q^1 + \Z \q^2$ is
identified as a sub-lattice of $\G_{m,1}$ by setting,
\beq\label{3.6}
 \q^1=\e^1+\e^2,\;\; \q^2=\e^1-\e^2 \ .
\eeq
Note that $\det G_L =4m=2^2 \det G_{m,1}$.
(This is a necessary condition for the lattice $L$ to
be a sub-lattice of index two of $\G_{m,1}$.)

(b) The vertex operator $E^g (z)$ is
constructed in the standard fashion from the
electric current $J(z)$ (see Appendix A) and satisfies the OPE:
\beq\label{3.2}
J(z_1)E^{g}(z_2)\sim \frac{g}{z_{12}}E^{g}(z_2)
\quad\Leftrightarrow\quad
[J(z_1),E^{g}(z_2)]=g\d(z_{12})E^{g}(z_2) \ .
\eeq
The second $\U1$ current (commuting with $J$) is the
counterpart of the ``spin current'' (\ref{1.8b}) in the unitary
$c=2$ theory:
\beq\label{3.3}
j(z)=\np{\psi^*(z)\psi(z)}\ , \qquad
[\psi^{(*)}(z_1),E^{g}(z_2)]=0=[j(z_1),J(z_2)]\ ,
\eeq
with both $j$ and $J$ normalized by
$\la 0|j(z_1)j(z_2)|0\ra=\la 0|J(z_1)J(z_2)|0\ra=z^{-2}_{12}$.

The  automorphism $\alpha[A]$,  $A \in\A_{m,1}$  is defined by:
\beq\label{3.4}
\alpha \left[A \right] = \ex^{i\pi J^1_0} A \ex^{-i\pi J^1_0} \ ,\qquad
J^1_0=\frac{1}{\sqrt m}J_0+j_0 \ ,
\eeq
(note that $ \alpha[\alpha[A]] = A$ for all $A\in\A_{m,1}$).
The property (\ref{action}) is implied by (\ref{3.4}) (in view of
(\ref{3.2}) and (\ref{3.3})).
The invariance of (\ref{3.5}) is then obvious.

{\bf Remark 2.1} This provides a simple example of the $\Z_2$
orbifold construction \cite{kt,dv3}; in the present case,
the orbifold actually
corresponds to a manifold (with no singular points) because both $\U1$
currents  are invariant under the parity operation.

\subsection{Superselection Sectors: Spectrum of Charges and Dimensions;
Partition Function}

We proceed to studying the {\it positive-energy representations}
of $\A(L)$ which define the {\it superselection sectors}
for our RCFT and are equipped with the fusion rules
corresponding to the addition of charges.

\begin{prop}
(a) The superselection sectors ${\cal H}_\lambda$
of the $\A(L)$ theory are labelled by the elements of the cyclic group,
\beq
\label{3.9}
L^*/L\simeq \Z_{4m} \ , \qquad (|L^*/L|=\det G_L=4m) \ ,
\eeq
where $L^*$ is the lattice dual to $L$:
\beq
\label{3.10}
L^*= \Z\q^*_1 + \Z\q^*_2 , \; (\q^i|\q^*_j)=\d^i_j \ ,\quad
G_{L^*}=\frac{1}{4m}\left(\matrix{
m+1 & 1-m \cr
1-m & m+1 \cr}\right) \ ,
\eeq
while $L^*/L =\{ \lambda \q^*_1 + L ; \lambda \ {\rm mod}\ 4m \}$.

(b) The ({\em visible}-in the terminology of ref. \cite{fst})
 electric charge vector $\Q$, whose square gives the filling
fraction of the model, belongs to the orthogonal sub-lattice
$\G^*_{m,1}=\G^*_m \oplus\G^*_1$ of $L^*$:
\beqa
\label{3.11}
\Q&=&\q^*_1+\q^*_2=\e^*_1,\;\; (\e^i|\e^*_j)=\d^i_j \nn
&& \Rightarrow (\Q|\q^i)=1 \ , \quad |\Q|^2=\nu=\frac{1}{m}\ .
\eeqa
The cyclic group (\ref{3.9}) is generated by either of the four
cosets $\pm\q^*_i+L$.

(c) The characters $\chi_\lambda (\tau,\z; m)$ of the
coset $(\lambda)$ is expressed in terms of sums of products of
$c=1$ lattice characters:
\beq
\label{3.13}
\chi_\l(\tau,\z;m)=
\ex^{-\frac{\pi}{m}\frac{(\I\z)^2}{\I\tau}} \mathrm{ch}_L^\l(\tau,\z),
\eeq
\beqa
\label{kform}
\mathrm{ch}_L^\l(\tau,\z) &\equiv &
{\rm tr}_{{\cal H}_\lambda} \left(
{\rm e}^{2\pi i \left[ \tau \left( L_0 -1/12 \right) +
\z J_0/\sqrt{m} \right] } \right) \nn
&=& K_\l(\tau;4)  K_\l(\tau,2\z;4m)+
K_{\l+2}(\tau;4)  K_{\l+2m}(\tau,2\z;4m) \ , \
\eeqa
where $K_l(\tau,\z;M)$ is given by,
\begin{alphalabel}\label{kfun}
\beq\label{kfuna}
K_l(\tau,\z;M)=\frac{1}{\eta(\tau)}\sum_{n\in\Z}
q^{\frac{M}{2}(n+\frac{l}{M})^2}\ex^{2\pi i \z(n+\frac{l}{M})}\ ,
\eeq
\beq\label{kfunb}
q=\ex^{2\pi i\tau}, \qquad
\eta(\tau)=q^{\frac{1}{24}}\prod_{n=1}^\infty (1-q^n),
\eeq
\end{alphalabel}
and we used the notation $ K_l(\tau;M)=K_l(\tau,0;M)$.
\end{prop}

{\bf Remark 2.2.} The non-analytic prefactor multiplying the character
${\rm ch}^{\lambda}_L$ corresponds to a constant term added to the Hamiltonian
and ensures the invariance under the ``spectral flow'' \cite{cz,gan}
of the resulting partition function (see Eq. (\ref{3.19b}) below).

Sketch of the proof. Statement (a) is known
(see Appendix A for a brief review of background material).
(b) One can choose in each class a representative
$\q^*$ such that the absolute value of the electric charge
$\vert ( \Q \vert \q^* ) \vert$ and its conformal dimension
$\Delta = \vert \q^* \vert^2 /2$ are minimal.
In Table \ref{tab.1}, we list the representatives of
each coset $(\lambda)=\lambda \q_1^* + L$ along with
the corresponding electric charge and conformal dimension.

The periodicity condition:
\beq\label{3.15}
K_{\l+M}(\tau,\z;M)= K_{\l}(\tau,\z;M) \ ,
\eeq
confirms that there are precisely $4m$ different characters $\chi_\l$
(\ref{kform}). We also note the symmetry property:
\beq\label{3.16}
K_{-\l}(\tau,-\z;M)= K_{\l}(\tau,\z;M) \ ,
\eeq
which clarifies, in particular, the observation that the cosets
$( \pm\l) $ in Table \ref{tab.1} give rise to the same conformal
dimension.

\begin{table}[H]
\caption{\label{tab.1}{\bf $(m+1)(m+1)(m-1) $ superselection sectors}}
\begin{center}
\begin{tabular}{|c||c|c|c|}
\hline
{\bf coset} & {\bf representative} & {\bf charge} &
{\bf dimension} \\
$(\lambda)=\lambda \q_1^* + L$ &
$\q^*$ & $ \left(\Q\vert \q^* \right)$ &
$ \Delta =\frac{1}{2}\left\vert\q^*\right\vert^2$ \\
\hline\hline
$(0)$ & ${\bf 0}$ & $0 $ & $0 $ \\
\hline
$(\pm 1)$ & $\pm \q_1^*$  &$\pm\frac{1}{2m}$ &
$\frac{m+1}{8m} $ \\
\hline
$\pm (2m-1) $ & $\pm \q^*_2 $ & $ \pm\frac{1}{2m}$ & $ \frac{m+1}{8m}$ \\
\hline
$\cdots $ & $\cdots $ & $ \cdots$ & $\cdots $ \\
\hline
$(\pm m) $ & $\pm (-1)^{m/2} \h\e^1$ & $ \pm (-1)^{m/2} \h$
& $ \frac{m}{8}$ \\
\hline
$(m\pm 1) $ &
$\h \left\{ \pm\left( m-1\right)\Q-\pm (-1)^{m/2} \h\e^2\right\} $ &
$\pm\frac{m-1}{4} $ & $\frac{m^2-m+1}{8m} $ \\
\hline
$\cdots $ & $\cdots $ & $\cdots $ & $ \cdots$ \\
\hline
$(2m) $ & $\e^2 $ & $0 $ & $\h $ \\
\hline
\end{tabular}
\end{center}
\end{table}

The partition function of any RCFT is given by
the quadratic form \cite{cft}:
\[
Z = \sum_{\lambda,\bar\lambda}
{\cal N}_{\lambda\bar\lambda} \
\chi_\lambda \ \overline\chi_{\bar\lambda} \ ,
\]
where the chiral (resp. antichiral) factors $\chi (\bar\chi)$
pertain to the outer
(inner) edge of the annular quantum Hall sample \cite{cz}.
The integer coefficients ${\cal N}_{\lambda\bar\lambda}$ are
obtained by the conditions of modular invariance of
$Z$ which are specific for the quantum
Hall systems. They have been formulated in Ref.\cite{cz} and
analyzed for lattice theories in Ref.\cite{gan};
$Z$ should be invariant under the transformations $S$ and $T^2$
which generate a subgroup of $ SL(2,\Z)$,
as well as under the $U$ and $V$ transformations specified below.

The SL(2,$\Z$) transformation properties of $K_\l$,
\beqa
T\ :\ K_{\l}(\tau+1,\z;M)&=&
\ex^{i\pi(\frac{\l^2}{M}-\frac{1}{12})} K_{\l}(\tau,\z;M) \ ,\nn
S\ :\ \ K_{\l}(-\frac{1}{\tau},\frac{\z}{\tau};M) &=&
\frac{1}{\sqrt{M}}\sum_{\mu\mod M}
\ex^{-2\pi i\frac{\l\mu}{M}}K_{\mu}(\tau,\z;M) \ ,
\eeqa
imply the following transformations for $\chi_\l$:
\begin{alphalabel}\label{3.17}
\beqa
T^2 : \chi_{\l}(\tau+2,\z)&=& \ex^{i\pi(\l^2\frac{m+1}{2m})-\frac{1}{3}}
\chi_{\l}(\tau,\z) \ , \label{3.17a} \\
S \ :\ 
\chi_{\l}(-\frac{1}{\tau},\frac{\z}{\tau})&=&\ex^{i\frac{\pi\z^2}{m\tau}}
\sum_{\mu\mod M} S_{\l\mu} \chi_{\mu}(\tau,\z) \ ,\nn
&&S_{\l\mu} = \frac{1}{2\sqrt{m}}\ex^{-i\pi\frac{m+1}{2m}\l\mu}\ .
\label{3.17b}\eeqa
\end{alphalabel}

The diagonal partition function:
\beq\label{3.18}
Z_{(m+1)(m+1)(m-1)}=\sum_{\l=1-2m}^{2m}\vert \chi_\l(\tau,\z)  \vert^2
\eeq
is then invariant under (\ref{3.17}) as well as under the $\z$-shift
$U$ and the spectral flow $V$:
\begin{alphalabel}\label{3.19}
\beqa
\label{3.19a}
U\ :\ \chi_\l(\tau,\z)& \rightarrow& \chi_\l(\tau,\z+1) =
\ex^{i\pi\frac{\l}{m}}  \chi_\l(\tau,\z) \ ,\\
\label{3.19b}
V\ :\ \chi_\l(\tau,\z) & \rightarrow &\chi_\l(\tau,\z+\tau) =
\ex^{\frac{-2\pi i}{m} Re(\z+\frac{\tau}{2})}  \chi_{\l+2m+2}(\tau,\z).
\eeqa
\end{alphalabel}
This diagonal partition function has the standard form for a lattice
RCFT discussed at length in Refs.\cite{cz}\cite{gan}.

Let us briefly recall here the physical meaning of these modular
invariance conditions \cite{cz}: the $T^2$ (resp. $U$) conditions
require that the observable electrons have half-integer spin 
(resp. integer charge); $S$ (resp. $V$) are self-consistency conditions
for the completeness of the excitations under the change of temperature
(resp. electric potential).

In the following, we shall consider the modular invariance of the
partition function as part of the building criteria; this leads to
the following postulate:

{\bf P1.}{\em The partition function of the RCFT describing a
fractional QH system should be invariant under (\ref{3.17}) and (\ref{3.19}).}

In the following, we are mostly interested in the 331 theory,
corresponding to $m=2$;
in Section $4$ and Appendix $C$, we prove that the partition function
(\ref{3.18}) is the unique solution to the modular invariance
conditions P1.

An explicit expression of the 331 partition function (\ref{3.18}), $m=2$,
is useful for the following discussion:
\beqa\label{z331}
Z_{331}&=& \sum_{r=0}^1 \Bigl\{
\vert K_0(\tau;4)K_{2r}(\tau,2\z;8 )+ K_2(\tau;4)K_{2r+4}(\tau,2\z;8)\vert^2\nn
&&+\vert  K_2(\tau;4)K_{2r}(\tau,2\z;8 )+
K_0(\tau;4)K_{2r+4}(\tau,2\z;8)  \vert^2   \nn
&&+\vert  K_1(\tau;4)K_{2r+1}(\tau,2\z;8 )+
K_{-1}(\tau;4)K_{2r-3}(\tau,2\z;8)\vert^2  \nn
&&+\vert  K_{-1}(\tau;4)K_{2r+1}(\tau,2\z;8 )+
K_1(\tau;4)K_{2r-3}(\tau,2\z;8)\vert^2 \Bigr\}\ ;
\eeqa
(we used the periodicity condition (\ref{3.15})).
This expression coincides, term by term, with that obtained in Ref.
\cite{mire}, for $\z=0$ (see their Eq. (4.20)), 
by taking into account the symmetry (\ref{3.16}) for the last two terms.

%-3-------------------------------------

\sectionnew{SU(2) Invariance versus Unitarity in the Haldane-Rezayi
Model: the Mapping from $c=-2$ to $c=1$}

%-3.1-----------------------------------

\subsection{SU(2) Covariant OPE of $\psi_\r(z)\psi_\s(w)$}

The four-point function of  the fermionic field $\psi_\r$,
$\r = \pm 1/2$, appearing in the HR wave function (\ref{1.5}) and
(\ref{1.6}) can be written in a manifestly SU(2) invariant form:
\beq\label{2.1}
\la 0| \psi_{\r_1}(z_1)\psi_{\r_2}(z_2)\psi_{\r_3}(z_3)\psi_{\r_4}(z_4) |0\ra=
\frac{\ep_{\r_1\r_2}\ep_{\r_3\r_4} }{z_{12}^2 z_{34}^2} +
\frac{\ep_{\r_1\r_4}\ep_{\r_2\r_3} }{z_{14}^2 z_{23}^2}-
\frac{\ep_{\r_1\r_3}\ep_{\r_2\r_4} }{z_{13}^2 z_{24}^2} \ .
\eeq
(Note that it is  only non-vanishing for $\r_1+\r_2+\r_3+\r_4=0$
 and then the right hand side of (\ref{2.1}) involves two non-zero terms.)
It follows that the OPE of two $\psi$ fields can also be written in an
SU(2) covariant form (see Appendix B).
We stress that this OPE just follows from the expression for the
$2N$-point correlation function: it does not use the Virasoro properties
of the fields; in fact, it does not require the knowledge of the
stress energy tensor (and admits different CFT implementations).
Keeping the first three terms in
the small distance expansion we can write (using (\ref{A.2}) and
(\ref{A.14})):
\beqa\label{2.2}
\psi_\r(z) \psi_\s(w)&=& \ep_{\r\s} \left \{
\frac{1}{(z-w)^2} - \T(w) - \frac{z-w}{2} \T'(w) \right\} \nn
&-&(z-w) V_{\r+\s}(w) + O((z-w)^2) \ ,
\eeqa
where $\T$ and $V$ are composite fields of $\psi_\r$,
\beqa\label{2.3}
\T(z) &=& \np{\psi_-(z) \psi_+(z)} \ , \nn
V_{\r+\s}(z) &= & \h\np{ \left\{\psi_\r(z)\partial \psi_\s(z)-
(\partial  \psi_\r(z)) \psi_\s(z)\right\} },
\eeqa
(cf. (\ref{1.11'})). They satisfy:
\beq\label{2.4}
\la 0| \psi_\r(z_1)  \psi_\s(z_2) \T(z_3)|0\ra=
\frac{\ep_{\r\s}}{z_{13}^2 z_{23}^2}\ ,\qquad
\la 0| \T(z_1)\T(z_2) |0\ra=-\frac{1}{z_{12}^4},
\eeq
\beq\label{2.5}
\la 0| \psi_\r(z_1)  \psi_\s(z_2) V_a(z_3)|0\ra=
-\frac{z_{12}}{z_{13}^3 z_{23}^3} G_{\r+\s,a}\ ,\quad
\la 0| V_a(z_1)V_b(z_2)|0\ra=\frac{G_{ab}^0}{z_{12}^6},
\eeq
where the non-zero elements of $G_{ab}$ (for $a,b=0,\pm 1$)
are $G_{00}=-1$ and $G_{1 -1}=G_{-1 1}=2$.
A salient feature of this OPE is the absence of the current $j$
(\ref{1.8b}) from the right hand side of (\ref{2.2}); however, it
will reenter our discussion when we impose unitarity.

%-3.2--------------------------------

\subsection{Other Choices of the Stress Tensor}

If we now assume that  $\psi_\r(z)$ are  primary fields
(of dimension 1), we should identify $\T$ with the stress tensor
of the theory; then, we arrive, in agreement with \cite{ww}, at a
non-unitary conformal model with $c=-2$ (indeed, one
identifies the coefficient $-1$ in front of
$\T$ in the right hand side of (\ref{2.2}) with $2\D/c$,
yielding $c=-2$ for $\D=\D(\pm 1)=1$ \cite{cft}).

One can, however,
preserve conformal invariance without allowing for negative norm
squares and negative dimensions. We remark that the
canonical anticommutation relations  (\ref{1.8a}) and the two-point
function (\ref{1.8a'},\ref{1.8b}) are sufficient for determining
the HR wave function (\ref{1.5}). These relations give room to
a family of stress tensors \cite{fms}:
\beq\label{2.6}
T_\kappa(z)=(1-\kappa)\np{\xi'(z)\eta(z)}-\kappa\np{\xi(z)\eta'(z)}=
\h\left[ \np{j^2(z)}+(1-2\kappa)j'(z)\right] \ ,\\
\eeq
\beq
j(z) = \np{\xi(z)\eta(z)}\ ,
\label{2.6'}\eeq
where  $T_0\equiv \T$.
The dimensions of $\xi(z)=\xi(z,\kappa)$ and $\eta(z)=\eta(z,1-\kappa)$
become $\kappa$ and $(1-\kappa)$, respectively. The current
anomaly and the Virasoro central charge depend on $\kappa$:
\begin{alphalabel}
\beq\label{2.7a}
T_\kappa(z_1) j(z_2) \sim  \frac{\partial}{\partial z_2}[z_{12}^{-1} j(z_2)]
+ (2\kappa-1) z_{12}^{-3} \ ,
\eeq
\beq\label{2.7b}
c_\kappa = 1-3(2\kappa-1)^2 \ .
\eeq
\end{alphalabel}
Clearly, $\psi_-=\xi'$ (\ref{1.9}) is only primary
for $\kappa=0$, i.e., $c=-2$, which accounts for this choice
in \cite{ww,mire,gfn}).

Here we remark that there is a {\em unique unitary point},
$\kappa=1/2$, $ c=1$, in which the current is primary, and we can
construct a model for
the HR state satisfying all desiderata listed in the Introduction.
In this case, we can identify $\xi$-$\eta$ with a pair of conjugate
Weyl spinors:
\begin{alphalabel} \label{2.8}
\beq\label{2.8a}
\eta(z,\h)\equiv\psi(z) \ ,\qquad \xi(z,\h)\equiv\psi^*(z) \ ;
\eeq
furthermore, the  stress  tensor (\ref{2.6}) takes the canonical form,
\beq\label{2.8b}
T(z) =T_{\h}(z) =\h \np{\left( {\psi^*}'(z) \psi(z)-
\psi^*(z) \psi'(z)\right)}   .
\eeq
\end{alphalabel}
The identification (\ref{2.8a}) implies a change of hermitean conjugation,
which is only possible at $\kappa=1/2$, where $\xi,\eta$ have the
same dimension. For $\kappa\neq 1/2$, we had instead
$\eta^*=\eta$ and $\xi^*=\xi$.
Note that the current $j(z)$ (\ref{2.6'}) is only hermitean with 
respect to the new conjugation rule (\ref{2.8a}).

The SU(2) invariant tensor $\T(z)$ appearing in the OPE (\ref{2.2})
becomes now non-hermitean: in terms of modes, setting
$\T(z)=\sum\limits_{n\in\Z} \Lc_n z^{-n-2}$ and
$T(z)=\sum\limits_{n\in\Z} L_n z^{-n-2}$, we have:
\beq\label{2.9}
\Lc_n=L_n-\frac{n+1}{2}j_n, \qquad \Lc^*_n=\Lc_{-n}-nj_{-n} \ ,\quad
L_n^*=L_{-n}\ ,\quad j_n^*=j_{-n}\ .
\eeq
The  tensor $\T(z)$ is an $SU(2)$ singlet, while the unitary
one, $T(z)$, is not, because $j_0 \propto S_3$ as discussed in Section
1.2.

{\bf Remark 3.1.} The inner products of $\widehat{u(1)}$ ground
states $|c,\s\ra$, where $\s$ is the eigenvalue of $j_0$,
 and the behaviour of $j_n$ under
conjugation depend on $c$. Indeed, the properties of the vacuum
ket $|-2,0\ra$  and of its bra counterpart are
dictated by the expression (\ref{1.8a}) for the $\xi$-$\eta$
two-point function and the mode expansions:
\beqa\label{2.25}
\xi(z)&=&\sum_{n}\xi_n z^{-n} \ ,\quad
\eta(w)=\sum_{n}\eta_n w^{-n-1} \ ,\nn
& \Rightarrow &
\eta_0\xi_0|-2,0\ra=|-2,0\ra \ ,\qquad \eta_0|-2,0\ra=0 \ .
\eeqa
Furthermore, for $c=-2$, the independent hermiticity of $\xi$ and $\eta$
implies the following conjugation properties of the current:
\beq\label{2.26}
j_n=\sum_{m\geq -n} \xi_{-m}\eta_{m+n}-
\sum_{m\geq 1} \eta_{-m}\xi_{m+n} \Rightarrow
j^*_n=\d_{n,0}-j_{-n} \ ,\qquad (c=-2).
\eeq
As a corollary, there are different bra-ket couplings in the two cases:
\beq\label{2.27}
\la -2,\s|-2,\tau\ra=\d^1_{\s+\tau} \ ,\qquad
\la 1,\s|1,\tau\ra=\d_{\s,\tau} .
\eeq
In particular, the bra counterpart of the ket vacuum $|-2,0\ra$
is $\la -2,1|$ both being annihilated by the $c=-2$ Virasoro
energy
\beq\label{2.28}
\Lc_0=\h j_0(j_0-1)+\sum_{n=1}^\infty j_{-n}j_n \ \ \left(=\Lc^*_0 \right).
\eeq
It is the presence of a second conjugation for $c=1$ (which,
incidently, does not change the hermiticity of $\Lc_0$) which
renders the $c=1$ model unitary.
It is remarkable that this $c$ dependence of inner products and
conjugation does not affect the correlation functions of charged fields.

{\bf Remark 3.2.} Lee and Wen \cite{lw} (building on a development
in \cite{mire}) propose a different way out of the unitarity problem.
They abandon the CFT calculations and give instead an explicit
construction of the Fermi fields in (\ref{1.6}):
\beqa
\psi_{\uparrow}(z)&=&\sum_{n=1}^\infty \sqrt{n}
(c_{n,\uparrow} z^{-n-1}+  c^*_{n,\downarrow} z^{n-1}) \ ,\label{LW1}\\
\psi_{\downarrow}(z)&=&\sum_{n=1}^\infty \sqrt{n}
(c_{n,\downarrow} z^{-n-1}-  c^*_{n,\uparrow} z^{n-1})\ , \nonumber
\eeqa
where $c_{n\r}$ satisfy canonical anticommutation relations,
which imply (\ref{1.7}) (for $\psi_\pm=\psi_{\uparrow\downarrow}$).
The Hamiltonian of the system is defined by the manifestly SU(2)
invariant expression:
\beq\label{LW3}
H= \sum_{n=1}^\infty n \left(c^*_{n,\uparrow} c_{n,\uparrow}+
   c^*_{n,\downarrow} c_{n,\downarrow} \right) \ ,
\eeq
(but no attempt is made to introduce a hermitean stress energy tensor).
$H$ and $\psi_\r$ are found to satisfy the standard commutation relations
for a dimension $1$ field:
$[H, \psi_\r(z)]=(z\frac{\dd}{\dd z}+1)\psi_\r(z) $.
(Incidently, the expression (\ref{LW1}) for a pair of Fermi fields of
integer dimension is a special case of a procedure proposed in Ref.
\cite{st}  to circumvent the spin-statistics theorem
in any number of space-time dimensions.)

The fields $\psi_\r$ (\ref{LW1}) - much like ours,
(\ref{1.9})(\ref{2.8})- are not conjugate to each other.
The SU(2) symmetry, however, has a more fundamental status
in \cite{lw}: the Hamiltonian (\ref{LW3}) is SU(2) invariant,
the fields  $\psi_\uparrow$ and  $\psi_\downarrow$ (\ref{LW1})
have the same dimension 1. What we regard, on the other hand,
as a shortcoming of this model displayed is its
{\em non-locality whenever the fields hermitean conjugate} to (\ref{LW1})
are included; indeed, while:
\[ [\psi_\uparrow(z_1), \psi_\downarrow(z_2) ]_+=
\sum_{n=1}^\infty n(z_1^{n-1}z_2^{-n-1}- z_1^{-n-1}z_2^{n-1})=
\d'(z_{12}) \ , \]
one has:
\[ [\psi_\uparrow(z_1), \psi^*_\uparrow(z_2) ]_+=
\sum_{n=1}^\infty n(z_1^{n-1}z_2^{-n-1}+ z_1^{-n-1}z_2^{n-1})=
\frac{1}{z^2_{12}}+ \frac{1}{z^2_{21}} \ . \]

%-3.3------------------------------------

\subsection{Operator Correspondence between the
\mbox{$c=-2$} and \mbox{$c=1$} Theories}

We relate the $c=-2$ and the $c=1$ models by identifying their
$\widehat{u(1)}$ currents. Noting that Eq. (\ref{2.6}) expresses,
for $\kappa=0$ and $1/2$, the corresponding stress energy
tensors $\T$ and $T$ in terms of $j$, we consider the $\widehat{u(1)}$
extension of the $c=-2$ Virasoro algebra Vir(-2) and compare it with
the corresponding known $\U1$ extension at $c=1$.
Next, the Vir(-2) primary field $\psi_-(z)=\xi'(z)$ satisfies
$[j(z),\xi'(w)]=\xi(w)\frac{\partial}{\partial w} \d(z-w)$; hence, it
isn't primary with respect to the current algebra. As a result, it
is not primary with respect to Vir(1) either. On the other hand, the
$\widehat{u(1)}$- primary fields for $c=-2$  are also
$\U1$ primary for  $c=1$.
\begin{prop}
Let $\phi_\s$ be a Vir(-2) primary field of $j_0$-``charge"
$\s$ and dimension $\D_\s^{(-2)}$.
Then it is $\widehat{u(1)}$- and, by implication,
Vir(1)-primary iff the dimension and charge are related by:
\beq\label{2.15}
\D_\s^{(-2)}=\h\s(\s-1) \ ;
\eeq
moreover,  $\phi_\s$ satisfies the Knizhnik--Zamolodchikov equation \cite{cft},
\beq\label{2.16}
\frac{d}{dz}\phi_\s=\np{j\phi_\s}\equiv\sum_{n=0}^\infty
(j_{-n-1}z^n \phi_\s +\phi_\s j_n z^{-n-1}) \ .
\eeq
The Vir(1) dimension, on the other hand, is symmetric under
charge conjugation:
\beq\label{2.16'}
\D_\s^{(1)}=\h\s^2 \ .
\eeq
\end{prop}
The {\bf proof} uses standard current algebra techniques 
(see, e.g., \cite{cft}).

We remark that the dimension (\ref{2.15}) of the $Vir(-2)$ fields
reproduces the Kac formula $\D_l$ (\ref{1.10}) for
$l=2\s -1$; however, we find that only a subset of the degenerate 
$Vir(-2)$ fields (forming $SU(2)$ multiplets) satisfy the charge-dimension 
relation. A complete set of $\widehat{u(1)}$-primary
fields (which are single-valued in the vacuum sector) is given by:
\beqa\label{2.17}
&&\phi_\s(z)=\frac{\xi(z)}{0!} \frac{\xi'(z)}{1!} \cdots
\frac{\xi^{(\s-1)}(z)}{(\s-1)!} \ , \nn
&& \phi_{-\s}(z)=\frac{\eta(z)}{0!} \frac{\eta'(z)}{1!} \cdots
\frac{\eta^{(\s-1)}(z)}{(\s-1)!} \ ,
\eeqa
Each $\U1$ primary
$\phi_{-\s}$  can be identified with the lowest spin-projection member
$\phi_{s,-s}$ of the SU(2) spin multiplet $\phi_{sm}$ , $2s=\s$,
which was discussed in the Introduction (see the text preceding
Eq.(\ref{1.13'})).  All other fields $\phi_{sm}$, $m\neq -s$,
are still Vir(-2) primary fields but are not
$\widehat{u(1)}$-primary (and not even quasi-primary with respect to
Vir(1)).

In order to construct an RCFT involving the series (\ref{2.17})
we consider the extended chiral algebra $\A(c,\sigma^2)$
for both $c=-2$ and $c=1$, which is generated by the $\s=\pm 2$ local
Bose fields:
\beq\label{2.18}
\phi_2(z)=\xi(z)\xi'(z) \ , \qquad
\phi_{-2}(z)=\eta(z)\eta'(z).
\eeq
It is the even part of the superalgebra $\mathcal{S}(c)$ of the $\xi$-$\eta$
pair. The bosonic algebra $\A(c,4)$ has 4 irreducible representations
(with energy bounded below) whose state spaces $\H_\s^{(c)}$ split
into infinite direct sums of irreducible $\widehat{u(1)}$ modules
$\V_\s$:
\beq\label{2.19}
\H_\s^{(c)}=\bigoplus_{n\in\Z}\V_{\s+2n}, \qquad
\s=0,\pm \h,1.
\eeq
The integer ``charges", $\s=0,1$, correspond to the Neveu-Schwarz
sector of $\mathcal{S}(c)$; $\s=\pm\h$ label the Ramond sector.

Eqs. (\ref{2.19}) and (\ref{2.9}) for $n=0$ allow to compute the
$c=-2$ characters in terms of the standard
$c=1$ lattice characters,
\beq\label{2.21'}
\chi^{(1)}_\s(\tau,\z)=\mathrm{tr}_{\H^{(1)}_\s}
\ex^{2\pi i \{ \tau(L_0-\frac{1}{24}) + \h\z j_0 \} }=K_{2\s}(\tau,\z;4) \ ,
\eeq
where $K_\l (\tau,\xi;M)$ is defined in (\ref{kfun}).

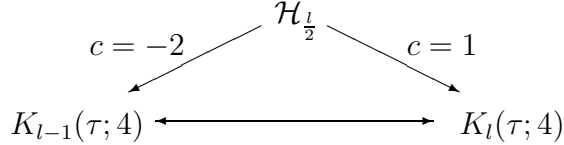
\begin{figure}[H]
\begin{center}
\begin{picture}(50,70)(20,-60)
\put(20,2){\makebox{$\mathcal{H}_{\frac{l}{2}}$}}
\put(-80,-40){\makebox{$K_{l-1}(\tau;4)$}}
\put(90,-40){\makebox{$K_{l}(\tau;4)$}}
\put(70,-10){\makebox{$c=1$}}
\put(-50,-10){\makebox{$c=-2$}}
\put(40,1){\vector(2,-1){50}}
\put(15,1){\vector(-2,-1){50}}
\put(10,-35){\vector(2,0){70}}
\put(10,-35){\vector(-2,0){35}}
\end{picture}
\caption{\label{graph} Shift of the characters in the mapping between the
$c=-2$ and $c=1$ theories.}
\end{center}
\end{figure}

Indeed, we have, in view of (\ref{2.9}),
\beqa\label{2.23}
\chi^{(-2)}_\s(\tau,\z)&=&\mathrm{tr}_{\H^{(-2)}_\s}
q^{\Lc_0+\frac{1}{12}} \ex^{2\pi i\z\h j_0}=
q^{\frac{1}{8}}\mathrm{tr}_{\H^{(1)}_\s}
q^{L_0-\frac{1}{24}}\ex^{2\pi i(\z-\tau)\h j_0} \nn
&=&q^{\frac{1}{8}}K_{2\s}(\tau,\z-\tau;4)=K_{2\s-1}(\tau,\z;4) \ .
\eeqa
This correspondence between the $c=-2$ and $c=1$
dimensions $\D_\s^{(c)}$ and characters
$\chi^{(c)}_\s$ is summarized in Table 2;
the ``index shift" for the characters
$\left( K_l(\tau;4)\rightarrow  K_{l+1}(\tau;4)\right)$  is
graphically represented on Fig. \ref{graph}.

Furthermore, in Ref. \cite{mire}, a partition function for the
$c=-2$ CFT has been proposed (see their Eq. (4.16)).
This can be written in our notation as follows:
\beqa\label{zHR}
Z_{HR}&=& \sum_{r=0}^1 \Bigl\{ 
\vert  K_0(\tau;4)K_{2r+1}(\tau,2\z;8 )+
K_2(\tau;4)K_{2r-3}(\tau,2\z;8)\vert^2 \nn
 &&+\vert  K_2(\tau;4)K_{2r+1}(\tau,2\z;8 )+
K_0(\tau;4)K_{2r-3}(\tau,2\z;8)\vert^2 \nn
&&+\vert  K_1(\tau;4)K_{2r}(\tau,2\z;8 )+
K_{-1}(\tau;4)K_{2r+4}(\tau,2\z;8)\vert^2 \nn
&&+\vert  K_{-1}(\tau;4)K_{2r}(\tau,2\z;8 )+
K_1(\tau;4)K_{2r+4}(\tau,2\z;8)\vert^2  \Bigr\} \ .
\eeqa
This expression is {\it not} modular invariant; however, 
under the $c=-2 \to 1$ mapping this remarkably becomes the 331 
partition function (\ref{z331}) found in Section 2:
\beq\label{map}
Z_{HR} \rightarrow Z_{331}\ ,\qquad {\rm for}\ \ c\to 1\ .
\eeq
This shows that the CFT for the HR state proposed in  Ref. \cite{mire},
once made unitary, becomes the same as the 331 CFT. Furthermore,
the parity rules coupling neutral and charged excitations is the same
in the two Hall fluids. Actually, the above mapping amounts to a
redefinition of the energy and of the scalar product on the same set of
states (see also Refs.\cite{lw,lud,ino}).
\begin{table}[H]
\caption{{\bf The $c=-2$ $\Longleftrightarrow$ $c=1$ correspondence}}
\begin{center}
\begin{tabular}{|r||c|c|c|c|c|} \hline
$l$ & $\s= l/2$ &
$\Delta_\s^{(c=-2)}= \s(\s-1)/2$ & $\chi^{(-2)}_\s(\tau)$
& $\Delta_\s^{(c=1)}= \s^2/2$  & $\chi^{(1)}_\s(\tau)$ \\ \hline \hline
$-1$ & $-1/2$ &  $ \; \ 3/8$  &  $K_2(\tau;4)$ & $1/8$   & $K_{-1}(\tau;4)$
\\ \hline
$0$ & $ \; \ 0$ & $\; \ 0$ &  $K_{-1}(\tau;4)$ &  $0$ & $K_{0}(\tau;4)$ \\
\hline
$1$ & $\; \ 1/2$ & $-1/8$ &  $K_0(\tau;4)$  & $1/8$ & $K_{1}(\tau;4)$ \\
\hline
$2$ & $\; \ 1$ & $\; \ 0$  & $K_{1}(\tau;4)$  & $1/2$& $K_{2}(\tau;4)$ \\
\hline
\end{tabular}
\end{center}
\end{table}

The correspondence between the dimensions in the two models is
given by:
\beq\label{2.20}
\D^{(-2)}_{\s+\h}+\frac{1}{12}=  \D^{(1)}_{\s}-\frac{1}{24} \ ,
\eeq
(the additive constant on each side being just $- c/24$).
The fields (\ref{2.17}), for integer $\s$, satisfy
(for both values of $c$) the {\em charge--statistics relation}:
\beq\label{2.21}
\phi_\s(z)\phi_\tau(w)=(-1)^{\s\tau} \phi_\tau(w)\phi_\s(z).
\eeq

%-4-------------------------------

\sectionnew{Admissible $c=2$ Descriptions of the $\nu=\h$ Double
Layer States}
In the previous section we have shown that the 331 model and the
(unitarized) HR model of Ref. \cite{mire} share the same partition function
and therefore have the same RCFT description. Nevertheless, we have
not so far excluded the possibility of other modular invariants made by the
same basis of states which could correspond to an alternative RCFT
for the HR theory.
The construction of the 331 partition function  in Section 2 shows that
its chiral building blocks (\ref{kform})  are expressed as sums of products
of $\A(\G_8)$- and $\A(\G_4)$- characters corresponding to the orthogonal
lattice
\beq
\label{4.1}
\G_{8,4}=\G_{8}\oplus \G_{4} \subset L \subset \G_{2,1} \subset \G^*_{2,1}
\subset L^* \subset  \G^*_{8,4}.
\eeq
This orthogonal basis of states allows, in principle, for other RCFTs
with different couplings between neutral and charged sectors, i.e. with
different parity rules. We have the following  result.
\begin{thm}
There are $7$ distinguishable RCFT models with ($c=2$ extensions of the)
chiral algebra $\A(\G_{8,4})$ that have modular invariant partition functions
(see  P1 in Section 2.2).  
Six of them factorize into products of neutral and charged parts
\beqa
Z_{8,4}^{(l)}(\tau,\z)&=& Z_{8}^{(l)}(\tau,\z) Z_{4}(\tau,0),\quad
l=1,-3, \label{4.2}\\
Z_{8,1}^{(l)}(\tau,\z)&=& Z_{8}^{(l)}(\tau,\z) Z_{1}(\tau,0),\quad l=1,-3, 
\label{4.4}\\
Z_{2,4}(\tau,\z)&=&Z_{2}(\tau,\z)Z_{4}(\tau,0) \ ,\label{4.5}\\
Z_{2,1}(\tau,\z)&=& Z_{2}(\tau,\z)Z_{1}(\tau,0) \ , \label{4.6}
\eeqa
where
\beq\label{4.3}
Z_{M}^{(l)}(\tau,\z)=\ex^{-\frac{4\pi}{M}\frac{(\I\z)^2}{\I\tau}}
\sum_{\l\mod M} K_\l(\tau,2\z;M)\overline{ K_{l \l}} (\tau,2\z;M) \ ,
\eeq
and $Z_M(\tau;\z)\equiv Z^{(1)}(\tau,\z)$.
The only non-factorizable one is $Z_{331}$ {\em (\ref{3.18})}.
\end{thm}
{\bf Remark 4.1.} Here and in what follows we deal exclusively with
{\em physical} partition functions in which all characters enter
with non-negative integer multiplicities and the vacuum character
appears with multiplicity one (cf. Refs.\cite{cz,gan}).

A proof of Theorem 4.1, based on ref. \cite{gan}, is given in Appendix C.

We shall analyze the seven RCFT singled out by the above theorem in order
to select those which can describe the $\nu=1/2$ state.
To begin with, we note that all 7 partition functions contain a pair of fields
with the properties of the electrons in the two layers,
\beq\label{4.7}
\psi^1(z)=\psi^*(z)E^{-\sqrt{2}}(z)\ ,\qquad 
\psi^2(w)=\psi(w)E^{-\sqrt{2}}(w) \ ,
\eeq
of charges $\q^i$ with Gram matrix and electric charge:
\beq\label{4.8}
\left(\matrix{
(\q^1|\q^1) & (\q^1|\q^2) \cr
(\q^2|\q^1) & (\q^2|\q^2) \cr}\right)=
\left(\matrix{
3 & 1 \cr
1 & 3 \cr}\right),\;\;\;
(\Q|\q^i)=1, \;\; i=1,2,\;\;\; |\Q|^2=\nu=\h
\eeq
(thus $\psi^i(z)$ can be identified with $E^{\q^i}(z)$ in
the notation of Appendix A). This has been made explicit in
Eq. (\ref{3.5}) for the $\Z_2$ orbifold model of partition function
$Z_{331}$ (\ref{z331}). Such fermion fields also belong
to the chiral algebra of the $Z_{2,1}$  model (Eq. (\ref{4.6}))
and to suitable extensions $\A=\A(\G_{\mathrm{phys}})$ of the lattice
chiral algebras $\A(\G)$,  $\G=\G_{2.4^{\mu},4^{\nu}},\; \mu,\nu=0,1,$
corresponding to the other six cases.
Here  $\G_{\mathrm{phys}}$ is an {\em integral extension } of $\G$
satisfying \cite{fst}:
\beqa\label{4.9}
\G&\subset&\G_{\mathrm{phys}}\subset\G^*,\nn
\exists\q^i&\in& \G_{\mathrm{phys}},\ i=1,2, \ \ \mathrm{so \ that}\ \ 
(\Q|\q^i)=1=|\q^i|^2 \ \mod 2.
\eeqa
In fact, if we exclude the case $\G_{\mathrm{phys}}=L\supset \G_{8,4}$
which  naturally leads to $Z_{331}$ for all 3 lattices
($\G=\G_{8,4},\G_{8,1},\G_{2,4}$),  $\G_{\mathrm{phys}}$
coincides with the maximal integral extension  $\G_{2,1}$ of
$\G$; this contains two charges $\e^1$ and   $\e^2$ of lengths 2 and  1,
respectively (corresponding to the fields $E^{\pm\sqrt{2}}(z)$ and
$\psi^{(*)}$ of (\ref{4.7})) and hence their sum and difference
$\q^{1,2}=\e^1\pm\e^2$ satisfy (\ref{4.8}).

There are, however, important distinctions between the
 $Z_{331}$ model and the models with factorizable partition functions
(\ref{4.2}-\ref{4.6}). The common factorizable extensions $\G_{2,1}$
of the three $\G$ involves a pair of Fermi fields $\psi^{(*)}$  of
dimension $1/2$ - smaller than the dimension $3/2$ of the basic
electron fields $\psi^i$. Moreover, all factorizable models contradict
a natural  postulate about the charge--statistics relation
\cite{ctz2}\cite{fst}.

{\bf P2.} {\em If $\Q$ is the electric charge vector (cf. (\ref{3.11})),
then   there exists a $\q'\in\G_{\mathrm{phys}}$ such that
$(\Q|\q')=1$ and for any $\q\in\G_{\mathrm{phys}}$  we must have:
\beq\label{4.10}
(\Q|\q)=|\q|^2 \ \mod 2 \qquad \left(\Q\in\G^*_{\mathrm{phys}} \right)\ ,
\eeq
(i.e., $\Q$ is an odd vector with respect to $\G_{\mathrm{phys}}$ in
the terminology of ref. {\em \cite{fst}}).}

This postulate states that there exists an electron excitation in the spectrum
and that all observable charged excitations are made  out of electrons.

Thus the two postulates P1 and P2 together with the requirement:
\beq\label{4.11}
\G_{8,4}\subset \G_{\mathrm{phys}} \subset \G_{2,1}=\{\Z\e^1 +\Z\e^2 \}\ ,
\quad \Q=\e^*_1 \ ,
\eeq
yield $\G_{\mathrm{phys}}=L$ (\ref{3.6}), i.e., a unique  lattice CFT with
partition function $Z_{331}$.

%-5------------------------------------

\sectionnew{Gauge Reductions}

%-5.1------------------------------------

\subsection{The Pfaffian State as a Projection of the 331 State
in the Low-Barrier Limit}

The $c=3/2$ {\em Pfaffian state} \cite{more} could describe a
double-layer sample in the particular limit in which the tunnelling
amplitude between the layers is large and the two species of
electrons become indistinguishable \cite{gww}.
This limit can be described in CFT by performing the following (generalized)
gauge reduction on the 331 CFT of distinguishable electrons.
Splitting the Weyl spinor $\psi$ into real and imaginary Majorana
components $\varphi_{1,2}$ ($\varphi_j^*(z)=\varphi_j(z) $):
\beq\label{5.16}
\psi^*(z)=\frac{1}{\sqrt{2}}\left(\varphi_1(z)+i  \varphi_2(z)\right)\ ,
\qquad \psi(z)=\frac{1}{\sqrt{2}}\left(\varphi_1(z)-i  \varphi_2(z)\right)\ ,
\eeq
we observe that the two layers just differ by the sign of
$\varphi_2$; thus, in the limit of low potential barrier
between the two layers, the imaginary part $\varphi_2$ of $\psi$
should be ``gauged away''.
This is achieved by the usual coset construction \cite{gko},
which consists first in decomposing the stress tensor in two
parts, as follows:
\beq\label{5.17}
T_{L}(z)=T_{Pf}(z)+\h\np{\varphi_2(z)\varphi_2'(z)}.
\eeq
The last term is then projected out, leading to a $c=3/2$ CFT.
The resulting theory can once more be viewed as a $\Z_2$
orbifold - this time the chiral algebra $\A_{Pf}$ appears as the
$\Z_2$ invariant part of the tensor product algebra
$\A(\G_2) \otimes \A_{Ising}(\varphi_1)$.
The Pfaffian ground state wave function is fully antisymmetric \cite{more}:
\beqa\label{5.18}
\Phi_{Pf}(z_1,\ldots,z_{2N})&=&
\la 0| \varphi_1(z_1) \varphi_1(z_2)\cdots \varphi_1(z_{2N}) |0 \ra
\Phi_m(z_1,\ldots,z_{2N}) \nn
&=&Pf\left(\frac{1}{z_{ij}}\right)\Phi_m(z_1,\ldots,z_{2N})
\eeqa
where $\Phi_m$ is the usual Laughlin factor (\ref{1.4}) and the Pfaffian is,
\beq\label{pfaff}
Pf\left(\frac{1}{z_{ij}}\right)=\frac{1}{2^N N!}\sum_{\s\in S_{2N}}
\sg\s\prod_{k=1}^N
\frac{1}{z_{\s(2k-1)}-z_{\s(2k)}}.
\eeq

The resulting CFT exhibits some interesting features and deserves a
special attention.
The total antisymmetry of the wave function signals the fact that
the electrons of the two layers are indeed indistinguishable.
The neutral part (\ref{5.18}) of the wave function appears
as the vacuum expectation value of a product of free Majorana fermions
which generate the Neveu-Schwarz sector of the $c=1/2$ Ising model \cite{cft}.
The Pfaffian model has topological order $6$.
The resulting representation spaces $\H_\l$ ($\l=0,\pm 1,\pm 2, 4$)
have lowest conformal weights:
\beq\label{5.20}
\D_0=0\ ,\quad \D_{\pm 1}=\frac{1}{8}\ , \quad \D_{\pm 2}=\frac{1}{4}\ , \quad
\D_4=\h \ ,
\eeq
and the same charges as the 331 model - see Table \ref{tab.1}.

The characters $\mathrm{ch}_\l$
are expressed, in parallel with (\ref{kform}), in terms of products of the
Ising characters $\chi_\nu$, $\nu=0,1,2$ and the characters
$K_l(\tau,2\zeta;8)$ of the 331 CFT:
\begin{alphalabel}\label{5.19}
\beqa
\mathrm{ch}_{2n}(\tau,\z)&=&\chi_0(\tau)K_{2n}(\tau,2\z;8)+
\chi_2(\tau)K_{2n+4}(\tau,2\z;8) \ ,\quad n=0,\pm 1,2,  \label{5.19a}
\qquad \quad\\
 & &\chi_{0}(\tau) \pm \chi_2 =
q^{-1/48} \prod\limits_{n=1}^\infty \left( 1 \pm q^{n-\h}\right)  \ ;\nn
\mathrm{ch}_{\pm 1}(\tau,\z) &=& \chi_1(\tau)  \left(K_{\pm 1}(\tau,2\z;8)+
K_{\mp 3}(\tau,2\z;8)\right)\ ,\label{5.19b} \\ &&\chi_1(\tau)=q^{1/24}
\prod\limits_{n=1}^\infty \left( 1+q^{n}\right). \nonumber
\eeqa
\end{alphalabel}

The modular invariant partition function
of the Pfaffian state is obtained by the diagonal quadratic form
in the basis (\ref{5.19}):
\beqa\label{zpf}
 Z_{Pf}&=& \sum_{r=0}^1 \Bigl\{ \vert  \chi_0(\tau)K_{2r}(\tau,2\z;8 )+
\chi_2(\tau)K_{2r+4}(\tau,2\z;8)\vert^2  \nn
&&+\vert  \chi_2(\tau)K_{2r}(\tau,2\z;8 )+
\chi_0(\tau) K_{2r+4}(\tau,2\z;8)\vert^2 \nn
&&+ \vert  \chi_1(\tau) \left( K_{2r+1}(\tau,2\z;8)
+K_{2r-3}(\tau,2\z;8)\right) \vert^2  \Bigr\}.
\eeqa
This coincides with the expression proposed in Ref. \cite{mire} (Eq.(4.13));
we have verified that it satisfies the modular conditions
(postulate {\bf P1}); in particular, the Pfaffian $S$ matrix,
\beq\label{pfaf}
S_{mn}=\h \sin \left( \frac{\pi}{4} \s_m\s_n \right)\ex^{-i\frac{\pi}{4}mn},
\qquad
\s_n=\frac{3-(-1)^n}{2} , \qquad  m,n=0, \pm 1, \pm 2, 4.
\eeq
has dimension $6\times 6$; this matches the topological order $6$, in
agreement with the general arguments of Ref.\cite{cz}.

It is instructive to see the gauge reduction from the 331 CFT  to the
Pfaffian in the  partition functions.
Using the Jacobi identity, the neutral characters $K_l(\tau;4)$
in the 331 partition function (\ref{z331})
can be written in fermionic Fock-space form:
\beqa\label{prod}
K_{0,2}(\tau;4)&=&\frac{q^{-1/24}}{2} \left(
\prod\limits_{n=1}^\infty \left( 1+q^{n-\h}\right)^2 \pm
\prod\limits_{n=1}^\infty \left( 1-q^{n-\h}\right)^2 \right) \ ,\nn
K_{\pm 1}(\tau;4)&=&q^{1/12}
\prod\limits_{n=1}^\infty \left( 1+q^{n}\right)^2 \ .
\eeqa
The projection of one Majorana fermion
can be realized  by eliminating the squares in these characters,
i.e.,
\beq
K_{0}(\tau;4) \  \rightarrow \ \chi_{0}(\tau) \ ,\quad
K_{2}(\tau;4) \ \rightarrow \ \chi_{2}(\tau) \  , \quad
K_{\pm 1}(\tau;4) \ \rightarrow \ \chi_1(\tau) \ .
\eeq
Moreover, only one for each pair of  twisted sectors of the 331 CFT
survives in (\ref{5.19}). After these transformations the 331 partition
function (\ref{z331}) is seen to become the Pfaffian one (\ref{zpf}).
It follows, in particular,  that the Pfaffian theory inherits the parity rule
of the 331 state.

The quantum dimension \cite{ver} of the representations ($\pm 1$) (of
characters (\ref{5.19b})) is $\sqrt{2}$ signalling the presence of a 
non-abelian (braid group) statistics for quasi-holes 
\cite{more,nw,rere}. The corresponding fusion rules \cite{cft} read:
\beq\label{5.21}
(\pm {\bf 1})\times (\pm {\bf 1})= ({\bf 2})+  ( -{\bf 2})\ , \qquad
({\bf 1})\times (- {\bf 1})= ({\bf 0})+  ( {\bf 4}).
\eeq

One may speculate that we observe here a manifestation of a more
general phenomenon: the non-abelian statistics comes from a gauge
reduction (in this case the projection $\varphi_2\rightarrow 0$)
of a lattice abelian model.
The U(1)$^n$ lattice models possess the characteristic $\W$ symmetry 
of the incompressible quantum Hall fluids under area preserving
diffeomorphisms \cite{ctz1}.
This has been found to be a crucial
feature of the CFT describing the single-layer hierarchical Hall states
\cite{ctz2}. Therefore, it seems natural to expect it to be
present in the paired Hall states as well.
However, the Pfaffian state is not described by a $\W$ CFT;
this is signalled by the fact that its central charge $c=3/2$ is not an
integer \cite{ctz2}. We thus see that the $\W$ symmetry is
broken by the gauge reduction, which is a quantum effect occuring in
the low potential barrier limit of a two-layer system.

It is important to remark that the $\W$ symmetry is still
present at the semiclassical level, which corresponds to the limit
$N\to\infty$ \cite{ctz1}; actually, in this limit, the Pfaffian
wave function (\ref{pfaff}) is dominated by the Laughlin factor
(angular momentum of order $O(N^2)$), 
while the Pfaffian is a subleading $O(1/N)$ relative correction.
The dominant term corresponds to the lattice RCFT which is
$\W$ symmetric.

%-5.2--------------------------------------

\subsection{Maximally Symmetric $c=3$ Description of Paired Hall States}

Until now we restricted our attention to rank $2$ charge lattices
which correspond to central charge $c=2$. It seems natural
to assume that the charge lattice associated with a QH state should have
minimal possible rank. From this point of view it may appear superfluous,
once we have a satisfactory description of the 331 state (\ref{1.13}) in
terms of the $c=2$ RCFT of Section 2, to search for higher rank lattices in
connection with this state.
It has been argued, however, by Fr\"{o}hlich et al.
(see Sections 5 and 7 and Table (B.2) in Appendix B of the second
of Ref. \cite{fst}), that a rank $3$ lattice provides a ``maximally symmetric"
RCFT for this $\nu=1/2$ state.  It involves an
$SU(2)\times SU(2)$ current algebra symmetry corresponding to spin
rotation and layer; it is realized by a pair of Weyl fermions which
are intertwined with the Laughlin boson.

We shall also demonstrate that the $c=2$ CFT considered
in the Sections 2 appears as a U(1) gauge reduction of this
one, by projecting out the difference of the two Weyl currents.
This identifies layer and spin quantum numbers of excitations,
yielding the 331 CFT.
In Ref.\cite{fst}, the chiral QH lattice, - i.e., the pair $(\G,\Q)$, is
defined in a {\em normal basis} as follows:
\beqa
\G=\G_{\mathrm{phys}}&=&\{ \Z\q+\Z\a^1+ \Z\a^2 \} \ ,\nn
G_\G &=& \left(\matrix{
3 & 1 & 1\cr
1 & 2 & 0 \cr
1 & 0 & 2 }\right), \qquad \Q=\q^*= {\q\over 2}-\frac{\a^1+\a^2}{4} \ .
\label{5.1}\eeqa
The structure of superselection sectors and their fusion rules
is similar to that of the 331 model. We have:
\beq\label{5.2}
\G^*/\G=\Z_8 \;\;\;(\Rightarrow |\G|=\det G_\G=8 ) \ ;
\eeq
this shows \cite{cz} that the topological order is again 8.
In Table~\ref{tab.3}, we display representatives of the non zero-elements of
the additive group $\G^*/\G$ (in parallel to the basis of $L^*/L$ given in
Table~\ref{tab.1}).

\begin{table}[H]
\caption{{\bf Superselection sectors for $\A(\G)$ }}
\label{tab.3}
\begin{center}
\begin{tabular}{|c||c|c|c|}
\hline
{\bf coset} & {\bf representative} & {\bf charge} &
{\bf dimension} \\
$(\lambda)=\lambda \a_1^* + \G$ &
$\q^*$ & $ \left(\Q\vert \q^* \right)$ &
$ \Delta =\frac{1}{2}\left\vert\q^*\right\vert^2$ \\
\hline\hline
$(0)$ & ${\bf 0}$ & $0 $ & $0 $ \\
\hline
$(\pm 1)$ & $\pm \a_1^*$  &$\mp\frac{1}{4}$ &
$\frac{5}{16} $ \\
\hline
$(\pm 2) $ & $\pm (2\a_1^* -\a^1) $ & $ \mp\frac{1}{2}$ & $ \frac{1}{4}$ \\
\hline
$(\pm 3) $ & $\pm ( 3\a^*_1-2\a^1 +\q)$ & $ \pm \frac{1}{4} $ &
$ \frac{5}{16}$ \\
\hline
$(4) $ & $\h (\a^1+\a^2) $ & $0 $ & $\frac{1}{2} $ \\
\hline
\end{tabular}
\end{center}
\end{table}

We note that the dimensions $\D_{2n}$ corresponding to the untwisted sector
of even charge cosets ($\D_0=0,\; \D_{\pm 2}= 1/4, \; \D_4=1/2$)
coincide precisely with those of the $\Z_4$ subgroup of $L^*/L$ given
in Table \ref{tab.1} for $m=2$ and $\lambda=0,\pm m, 2m$.
The same is true for the minimal (absolute) values of the inner
products of $\Q$ with vectors in each such coset, listed in the third
column of Table \ref{tab.3}. In both cases, {\em the smallest positive }
(fractional) {\em electric charge of a quasi-particle is $1/4$}.
The lattice $\G$ (\ref{5.1}) obeys the  postulate P2 and the
electric charge satisfies (\ref{4.8})  ($\vert \Q \vert^2 = \nu=1/2$).

In order to verify this statement we pass from the normal basis
(\ref{5.1}) of $\G$ to its {\em symmetric basis} \cite{fst}:
\begin{alphalabel}\label{5.4}
\beq\label{5.4a}
\q^1=\q-\a^1,\ \q^2=\q-\a^2,\ \q^3=\q, \ (\Rightarrow\ 
\a^i=\q^3-\q^i,\ i=1,2),
\eeq
characterized by,
\beq\label{5.4b}
G_\G(\q^i)=\left( (\q^i|\q^j)_{i,j=1,2,3} \right)=
\left(\matrix{
3 & 1 & 2\cr
1 & 3 & 2 \cr
2 & 2 & 3 }\right),\;\; (\Q|\q^i)=1.
\eeq
\end{alphalabel}
The resulting RCFT with chiral algebra $\A(\G)$ can  again be
viewed
as a $\Z_2$ orbifold of a tensor product algebra corresponding to the
orthogonal integral lattice $\G_{2,1,1}$ (that extends $\G$):
\beqa\label{5.5}
(\G\subset) \G_{2,1,1}&=&\{\Z\e^0\oplus\Z\e^1 \oplus\Z\e^2 \},\;\;\;
\e^0=2\Q=\q-\e^1,\; e^1=\h(\a^1+\a^2),\nn
&&e^2=\h(\a^1-\a^2) \;
(\e^{\mu}|\e^{\nu})=|\e^\mu|^2 \d_{\mu\nu},\;\mu,\nu=0,1,2, \nn
&&|\e^0|^2=2,\;\; |\e^1|^2 = |\e^2|^2 =1.
\eeqa
Indeed, if we introduce the current:
\beq\label{5.6}
J(z)= J^{\Q}(z) +  J^{\a^1}(z) \ ,\quad \a^1=\e^1+\e^2 \ ,\quad
J^{\Q}(z)=\h J^{\e^0}(z)\ ,
\eeq
then the inner automorphism $\alpha$ of the extended chiral algebra
given by
\beq\label{5.7}
\alpha(A)=\ex^{i\pi J_0}\ A \ \ex^{-i\pi J_0},\;\;\;
\mathrm{for} \;\;A\in\A_{2,1,1}\ ,
\eeq
leaves invariant each element of the physical subalgebra
$\A(\G)$ while changing
the sign of the basic charge shift operators $E^{\pm\e^\nu}$ and
the associated Bose (for $\nu=0$) and Fermi  (for $\nu=1,2$) local
charged fields (see Eqs. (\ref{B.9}-\ref{B.11}) of Appendix A).

The characters $\chi^\G_\l(\tau,\z)$ of the irreducible representations
of the chiral algebra $\A(\G)$ are given by the following counterpart
of (\ref{3.13})  and (\ref{kform})
\begin{alphalabel}\label{5.8}
\beq\label{5.8a}
\chi^\G_\l(\tau,\z) =\ex^{-\frac{\pi}{2}\frac{(\I\z)^2}{\I\tau}}
\mathrm{ch}_\G^\l(\tau,\z) \ ,
\eeq
\beq\label{5.8b}
\mathrm{ch}_\G^\l(\tau,\z)=K_0(\tau;2)  K_\l(\tau;2) K_{-\l}(\tau,2\z;8)+
K_1(\tau;2)K_{\l+1}(\tau;2)  K_{4-\l}(\tau,2\z;8)
\eeq
\end{alphalabel}
($\l\mod 8$) where the $K$-functions are again given by (\ref{kfun}).
 The proof of
(\ref{5.8}) is completely analogous to that of (\ref{kfun}) given
in Appendix A.

We end up once more with a diagonal modular invariant
partition function of type (\ref{3.18}),
\beq\label{5.9}
Z_\G(\tau,\z)=\sum_{\l=-3}^4 |\chi^\G_\l(\tau,\z)|^2.
\eeq
In addition, there is an $\widehat{su(2)}\otimes\widehat{su(2)} $ current
 subalgebra of $\A(\G)$ generated by the charged currents
$E^{\pm\a^i}(z)$ satisfying:
\beq\label{5.10}
[ E^{\a^i}(z_1),  E^{-\a^j}(z_2)]=
\left(J^{\a^i}(z_2) \d(z_{12})-\d'(z_{12})\right)\d_{ij}\ ,
\eeq
where $J^{\a}$ are the corresponding Cartan currents
 which commute  with the electromagnetic
(u(1) - ) current  $J^{\Q}(z)$ (defined in (\ref{5.6})). It provides
a local realization of the
$\mathrm{SU(2)}_{\mathrm{spin}}\times \mathrm{SU(2)}_\mathrm{layer}$
symmetry of the model (justifying the term ``maximally symmetric" of
\cite{fst}).

The rank $2$ $(\A(L))$ realization of the $\nu=\h$ state, given in the
preceding sections, is recovered if we gauge the u(1) current:
\beq\label{5.11}
I(z)(\equiv J^{\e^2}(z))=\h\left(J^{\a^1}(z)-J^{\a^2}(z) \right) \ ,
\eeq
generated by the pair of conjugate Weyl spinors,
\beq\label{5.12}
\varphi(z)=E^{-\e^2}(z)\ ,\quad \varphi^*(z)=E^{\e^2}(z)\ \Rightarrow\ 
I(z)=\np{\varphi^*(z)\varphi(z)} \ .
\eeq
In particular, the new stress tensor is:
\beq\label{5.13}
T_L(z)=\h\np{\Bigl\{ J_1(z)J^1(z) + J_2(z)J^2(z)\Bigr\}}
\eeq
where $\{J^i \}$ and $\{J_i \}$  are dual bases in the Cartan
subalgebra of $\A(L)$ (cf. (\ref{B.14})). 
$T_L(z)$ is obtained from the stress 
tensor $T_\G(z)$ of  $\A(\G)$ by subtracting the Sugawara contribution
of $I(z)$:
\beq\label{5.14}
T_L(z)=T_\G(z)-\h\np{I^2(z)}.
\eeq
In other words, $\A(L)$  is obtained from $\A(\G)$ by gauging out the
second factor in the extended algebra,
\beq\label{5.15}
\A(\G_{2,1,1})= \A(\G_{2,1})\otimes  \A(\G_{1}) \ ,
\eeq
corresponding to the lattice (\ref{5.5}) and then taking  the
$\Z_2$ orbifold.

Let us finally remark that the choice between this $c=3$
models and its 331 reduction should be decided by  experiment.

%-6------------------------------------------

\sectionnew{Conclusions}

We presented a unified description of the double-layer
$\nu=1/2$ Hall states which was based on two assumptions: (P1) the use
of lattice $c=2$ CFTs with modular invariant partition functions; and (P2) the
usual charge -- statistics relation for the observed electron-like excitations.

Our analysis has singled out a unique $c=2$ RCFT  which corresponds to the
already proposed model for the 331 state.  We have shown that  the $c=-2$  CFT
previously proposed for the HR state is in fact equivalent to the 331 CFT,
which is the unitary description of the same set of states. The result has
been obtained by a one-to-one mapping between the $c=-2$ and $c=1$ CFTs
(Section 3).  Furthermore, in Section 4 we proved that there are no alternative
lattice RCFTs for the HR state which are made of the same chiral
building blocks (e.g., there are no alternative parity rules for 
combining the fermion number and the charge).

The Pfaffian state has also been related to this  unique CFT by a gauge
reduction which has physical interpretation  in the double-layer
geometry. This projection eliminates  the fermionic U(1) current
and breaks the $\W$ symmetry at the quantum level; the parity rule for the
excitations remains the same.

In our description, the d-wave spin-singlet part of the HR wave function
(\ref{1.5}) is represented  in terms of  the $c=1$ Weyl fermion as follows:
\beqa\label{7.1}
\Psi_{ds} &=& \la 0|\partial \psi^*(z_1) \cdots
\partial\psi^*(z_N)\psi(w_1)\cdots \psi(w_N)  |0\ra \nn
&=& \frac{\partial^N}{\partial z_1 \cdots \partial z_N}
\det\left( \frac{1}{z_i-w_j}\right) = (-1)^N
\det\left( \frac{1}{\left(z_i-w_j\right)^2} \right)  \ .
\eeqa
The physical electron fields can also be written:
\beqa\label{7.2}
\np{j(z)\psi^1(z)}&=&\np{j(z)\psi^*(z)}E^{-\sqrt{2}}(z) \ ,\nn
\psi^2(z)&=&\psi(z)E^{-\sqrt{2}}(z) \ ;
\eeqa
here, we have used the K-Z equation \cite{cft} to substitute  the derivative
of $\psi^*$   by its product with the current $j$, and 
$\psi^1$ and $\psi^2$ are the electron fields of
the 331 state in Section 2.

Equations (\ref{7.1},\ref{7.2}) show that in our unitary description 
the HR ground state is an excited state of the 331 CFT.
Moreover, it has the same topological order
8 (for $m=2$, i.e.,  $\nu=1/2$) and its excitations obey abelian 
statistics as in any lattice RCFT.
These results are conclusive in the framework of lattice RCFTs.

On the  other hand, there are arguments in the literature in 
favour of distinguishing the HR from
the 331 state:
\begin{itemize}
\item{There is a wide spread opinion that the HR quasi-particles 
obey non-abelian statistics  \cite{ww,mire,lw,rere}. }
\item{The wave functions for the excitations of $2n$ quasi-holes on the sphere
are represented by symmetric polynomials spanning a $2^{2n-1}$ dimensional
space for the 331 model and a   $2^{2n-3}$  dimensional one for the HR state
 \cite{rere}. }
\item{The topological order of the HR state has been found to be 10 by
constructing its ground-state wave functions on the torus \cite{hr,wkk,rere} 
and by numerically diagonalizing the energy spectrum \cite{hr}.}
\end{itemize}

Finally, the unitary description of the HR state in this paper
does not explain the stability of the HR
ground state, which can naively decay in the 331 ground state.
Possible solutions to these problems might be found by describing
the HR with other, non-lattice conformal theories; 
for example, orbifold theories obtained from
the 331 RCFT, which give rise to non-abelian statistics.
Furthermore, other quantizations of the original $\xi$-$\eta$ system
might be possible based on a careful analysis of the zero modes,
as suggested in Refs.\cite{fms,gfn}.
These issues are left for future investigations.
\\
\vspace{0.5cm}
\\
\noindent
{\bf \Large Acknowledgements}

L.G. and I.T. would like to thank INFN, Sezione di Firenze, and SISSA, Trieste
for hospitality. Their work was supported in part by the Bulgarian National
Foundation for Scientific Research under contract F-404.
A.C. also thanks SISSA for hospitality and acknowledges the European
Community programme FMRX-CT96-0012. 
All three authors thank the Schr\"{o}dinger Institute for
Mathematical Physics, Vienna, for hospitality during 
the final stage of this work.

\appendix[Appendix ]

%-A-----------------------------------

\sectionnew{Charge Lattices, Cyclic Groups and Orbifolds}

We briefly recall some background material about integral
lattices (concise, physicist oriented surveys can be found
in \cite{fst,gan}).
An {\em Euclidean integral lattice} $\G$ {\em of rank $r$} is an
$r$-dimensional $\Z$-module (free abelian group) with a positive definite
integer valued bilinear form:
\beq\label{B.1}
\G=\Z\e^1+\cdots \Z\e^r,\;\;\; (\e^i|\e^j)=(\e^j|\e^i)\in\Z,
\eeq
\beq\label{B.2}
\G\ni \v \Rightarrow (\v|\v)\in\Z_+,
(\v|\v) =0 \Leftrightarrow \v=0.
\eeq
The basis $\{\e^1,\ldots ,\e^r \}$ of $\G$ is  determined up
to  GL(n,$\Z$) transformations,
\beq\label{B.3}
GL(n,\Z) \ni A=(a^i_j):\e^i\rightarrow  a^i_j\e^j,\;\;\;
a^i_j \in\Z,\;\; i,j=1,\ldots, r,\;\; \det A=\pm 1.
\eeq
Hence, the {\em determinant} $|\G|$ of the {\em Gram matrix } $G_\G$
is an invariant of the lattice $\G$. The {\em dual lattice} $\G^*$
is defined as the set of all vectors $\u\in\Rset^r$ such that
$(\u|\v)\in\Z$ for any $\v\in\G$.
The basis $\{\e^*_i \}\subset\G^*$ is dual to $\{\e^i \}\subset\G$
if $(\e^i|\e^*_j)=\d^i_j$. Clearly, $\G\subset\G^*$ is an (invariant)
subgroup of the (abelian) group $\G^*$. The quotient $\G^*/\G$ is a
finite abelian group (a product of cyclic groups) of order $|\G|$:
\beq\label{B.4}
|\G^*/\G|=\left(\frac{\det G_\G}{\det G_{\G^*}}\right)^\h=\det G_\G=|\G|
\;\;\; (G_{\G^*}=G_{\G}^{-1}).
\eeq
Let $L$ be a sub-lattice of the integral lattice $\G$. Then
$\G^*\subset L^*$ and the finite abelian groups $\G / L$ and
$L^* /\G^*$ are isomorphic:
\beq\label{B.5}
L\subset\G\subset\G^*\subset L^*, \;\;\;
L^*/\G^*\simeq \G/L.
\eeq

To each integral lattice $\G$ there corresponds a chiral vertex algebra
$\A(\G)$ (see, e.g., Section 1.2 of \cite{kt} and references therein).
It involves, to begin with, $r$ linearly independent u(1) currents
$J^i(z)=\sum_n J^i_n z^{-n-1}$ (in the basis $\{\e^i,\; i=1,\ldots,r \}$
of $\G$) such that:
\beq\label{B.7}
[J^i(z_1), J^j(z_2)]=-(\e^i|\e^j)\d'(z_{12}) \;\; \Leftrightarrow \;\;
[J^i_n, J^j_m]=(\e^i|\e^j) \d^0_{n+m}.
\eeq
It defines a free Bose field subalgebra $\A_r$ of $\A(\G)$.
We consider a (reducible) positive energy {\em vacuum representation}
of $\A_r$ in a Hilbert space $\H_\G$ that splits into an infinite direct
sum of irreducible $\A_r$ modules  $\H_{\v}$ with
distinguished cyclic vectors $|\v\ra$
$(\v\in\G)$:
\beq\label{B.8}
\H_\G=\bigoplus\limits_{\v\in\G} \H_{\v} ,\;\;
|\v\ra  \in   \H_{\v} ,\;\;\;
J^i_n |\v\ra  = (\e^i|\v) \d^0_{n}|\v\ra
\;\;\mathrm{for}\;\; n\geq 0.
\eeq
To each $(\u\in\G)$ we associate a unitary {\em shift operator}
$E^{\u}$ acting in $\H_\G$  such that:
\beq\label{B.9}
 E^{\u} |\v\ra =\ep(\u,\v)
 |\u+\v\ra \ ,\quad  J^i_n E^{\u}  =
E^{\u} \left(J^i_n+(\e^i|\u)\d^0_{n}\right).
\eeq
The factor $\ep(\u,\v)$    takes values $\pm 1$.
It is a 2-cocycle, - i.e.,
\[ \ep(\u_1,\u_2)\ep(\u_1+\u_2,
\u_3)=
\ep(\u_1,\u_2+\u_3)
\ep(\u_2,\u_3),
\] and satisfies
\[ \ep(\u,0)=\ep(0,\u)=1, \;\;\;
\ep(\u,\v)=
(-1)^{|\u|^2 |\v|^2-(\u|\v)}
\ep(\v,\u),\;\;\;
|\u|^2=(\u|\u),
\]
thus guaranteeing the normal spin-statistics relation for
different fields.

To each vector $\v=v_i\e^i$ in the real $r$-dimensional
space $\Rset^r$ we associate a u(1) current $J^{\v}$ by:
\beq\label{B.10}
J^{\v}(z)=v_i J^i(z) \ , \qquad ( J^i(z)= J^{\e^i}(z)).
\eeq
If  $\v\in\G$ then there exists a {\em local (Bose or Fermi)}
charged  field  $E^{\v}(z)$ such that:
\beq\label{B.11}
E^{\v}(z) |0\ra=\exp \left\{\sum_{n=1}^\infty
J_{-n}^{\v} \frac{z^n}{n} \right\}|\v\ra,\;\;\;
|\v\ra=E^{\v}|0\ra\ ,
\eeq
and
\beq\label{B.12}
E^{\u}(z)  E^{\v}(w)=
(-1)^{|\u|^2|\v|^2}
E^{\v}(w)   E^{\u}(z)\;\;\; \mathrm{for}\;\; w\neq z.
\eeq

Let $\{ J_i(z):= J^{\e^*_i}(z)\}$ be a basis of u(1) currents
dual to $\{ J^i(z)\}$. The stress energy tensor $T$ for each chiral
algebra $\A(\G)$ belongs to its subalgebra $\A_r$ and is given by
the Sugawara formula:
\beq\label{B.14}
T(z)=\h\np{J_i(z)J^i(z)} \ .
\eeq
It implies that the energy of a ground state vector $\v$
of $\A_r$  is given by:
\beq\label{B.15}
(L_0-\h |\v|^2)|\v\ra=0\;\;\; \mathrm{if}\;\;\;
(J_{in}-\d^0_n v_i)|\v\ra=0 \;\; \mathrm{for}\;\; n\geq 0, \ 
v_i=(\v|\e^*_i).
\eeq
The irreducible positive energy representations of $\A(\G)$ are
labeled by the elements of the finite abelian group $\G^*/\G$.
Let $\v^*\in\G^*$ be a representative of the coset
$\v^*+\G$ satisfying:
\beq\label{B.16}
|\v^*|^2=\mathrm{inf}_{\u\in\G}
\; |\v^*+\u|^2.
\eeq
Such a $\v^*$ is, in general, not unique in
$\v^* +\G$, however the representation space,
\beq\label{B.17}
\H_{\v^*+\G}=\bigoplus_{\u\in\G}
\H_{\v^*+\u}\ ,
\eeq
which  generalizes the vacuum space (\ref{B.8}) is clearly
independent of the choice of  $\v^*$.
The minimal (ground state)
energy in  $\H_{\v^*}$ is $\h |\v^* |^2$
(provided (\ref{B.16}) takes place).

Returning to the chain of embedded lattices
(\ref{B.5})  we shall demonstrate  that $\A(L)\subset \A(\G)$  appears
as an orbifold of $\G$.
\begin{prop}
The finite abelian group $L^*/\G^*$ acts by automorphisms on the
chiral algebra $\A(\G)$ leaving each element of its subalgebra
$\A_r\subset\A(\G)$ invariant.
\end{prop}
{\bf Proof.}
Indeed, for each $\u^*\in L^*$ we can define a
{\em gauge operator} $U(=U_{\u^*})=\ex^{2\pi i J_0^{\u^*}}$ satisfying:
\beq\label{B.18}
\alpha_{\u^*}[E^{\v}]= U E^{\v} U^{-1}=
\exp \left\{2\pi i (\u^*|\v) \right\} E^{\v} \ ,
\qquad (\v\in\G).
\eeq
(Although the operator $U$ depends on the choice of representative
$\u^*$ in the coset  $\u^*+\G^*$ the automorphism
$\alpha_{\u^*}$ of $\A(\G)$ does not.)
For $\v\in L (\subset \G)$  $(\u^*|\v)\in\Z$
so that $E^{\v}$ is unaltered by the automorphism
$\alpha_{\u^*}$ in (\ref{B.18}).

We are now prepared to apply Proposition A1 to the pair
$L\subset\G_{m,1}\equiv \G_m\oplus\G_1$ of integral lattices
studied in Section 2. In this case:
\beq\label{B.19}
\frac{|L|}{|\G_{m,1}|}=\frac{4m}{m}=2^2 \;\;\Rightarrow
L^*/\G^*_{m,1}\simeq\Z_2 \;\;\;
(\G^*_{m,1}=\G^*_{m}\oplus\G^*_{1}) \ ,
\eeq
and the non-trivial automorphism $\alpha$ is given by (\ref{action});
note that the coset $L^*/\G^*_{m,1}$  can be represented
by either of the four vectors $\pm \q^*_i$ and we have,
\beq\label{B.20}
J^{\q^*_1}(z)=\h J^1(z)=\h\left\{ \frac{1}{\sqrt{m}} J(z)+j(z)\right\}\ ,
\eeq
in the notation (\ref{3.2}-\ref{3.4}).

We shall now sketch a derivation of the formula (\ref{kform}) for the
character of the representation $ (\l)=\l\q^*_1+L$ of the orbifold
algebra $\A(L)$.
\begin{prop}
The character $\mathrm{ch}_L^\l(\tau,\z)$ {\em (\ref{kform})} is the chiral
partition function of the conformal Hamiltonian $L_0-\frac{2}{24}$ and
the visible charge operator $J^{\Q}_0$ in the $\A(L)$ module
$\H_\l \ (=\H_{\l\q^*_1+L}) $:
\beq\label{B.21}
\mathrm{ch}_L^\l(\tau,\z)=\mathrm{tr}_{\H_\l}\left\{
q^{L_0-\frac{1}{12}} \ex^{2\pi i \z J^{\Q}_0} \right\}
,\;\;\;
q=\ex^{2\pi i \tau}\ .
\eeq
\end{prop}
{\bf Proof.}
Substitute $\H_\l $ by the  direct sum (\ref{B.17}),
\beqa\label{B.22}
&&\H_\l=\bigoplus\limits_{n_1,n_2\in \Z}
\H_{\l\q^*_1+n_1\q^1+n_2\q^2}, \nn
&&n_1\q^1+n_2\q^2=(n_1+n_2)\e^1+ (n_1-n_2)\e^2.
\eeqa
The contribution of each term to the trace
(the sum over the currents' excitations in $\H_{\l}$ )
gives a factor
$[\eta(\tau)]^{-2} \ex^{2\pi i \{\frac{\tau}{2}(\u+\l\q^*_1)^2+
\z(\Q|\u + \l\q^*_1)\}}$
where $\Q$ is given by (\ref{3.11}). Noting that $n_1+n_2$ and $n_1-n_2$
have the same parity we split the resulting double sum in two terms:
one with $n_1+n_2=2n$,  $n_1-n_2=2n'$ and another with
$n_1+n_2=2n+1$,  $n_1-n_2=2n'+1$ . The first of these terms gives the first
summand in (\ref{kform}):
\[\frac{1}{\eta^2(\tau)}\sum_{n,n'}
q^{2m(n+\frac{\l}{4m})^2+ 2(n'+\frac{\l}{4})^2 }
\ex^{4\pi i \z(n+\frac{\l}{4m})} = K_\l(\tau;4)    K_\l(\tau,2\z;4m).
\]
Similarly, the second one reproduces the second term,
$K_{\l+2}(\tau;4)  K_{\l+2m}(\tau,2\z;4m)$,  thus completing the proof
of the proposition.

A different method of computing $\mathrm{ch}_L^\l(\tau,\z)$ (and hence of proving
Proposition A2) using technique of orbifold theory \cite{kt} is provided
in \cite{gt}.

%-B----------------------

\sectionnew{Conformal OPE for the HR Anticommuting Fields}

Standard (global) conformal OPE are written as series of
integrals of quasi-primary fields with respect to a given stress
energy tensor  ( see, e.g., Appendix A of
\cite{fpt}, \cite{bgt} and \cite{gt}). Here we shall  write down
such an expansion for the product of (free) fields
$\psi_\r(z)\psi_\s(w)$ without committing ourselves to a choice
of the stress energy tensor (and the associated Virasoro algebra). All
we shall need is the 4-point function (\ref{2.1}) with typical element:
\beqa\label{A.1}
&&\la 0| \psi_{+}(z_1)\psi_{-}(w_1)\psi_{+}(z_2)\psi_{-}(w_2) |0\ra=\nn
&&=(z_1-w_1)^{-2} (z_2-w_2)^{-2} - (z_1-w_2)^{-2} (w_1-z_2)^{-2}.
\eeqa
We shall verify (using techniques developed in the above references)
that it gives rise to an OPE of the form:
\beqa\label{A.2}
(z-w)^2 \psi_\r(z)\psi_\s(w)&=& \ep_{\r\s} \left\{
1- \int\limits_{w}^z [6\frac{(z-\z)(\z-w)}{z-w} \T(\z) \right. \nn
&&\left. +70 \frac{(z-\z)^3(\z-w)^3}{(z-w)^3} \L(\z) ] \dd\z +
(z-w)^6 S(z,w) \right\}  \nn
&&-30 \int\limits_{w}^z \frac{(z-\z)^2(\z-w)^2}{(z-w)^2}
V_{\r+\s}(\z) \dd\z \nn
&& + (z-w)^5 R_{\r+\s}(z,w)\ , \qquad \r,\s=\pm \h \ .
\eeqa
Here the $\T,\; \L$ and $V_a$  are (translation invariant) local
fields satisfying:
\begin{alphalabel}\label{A.3}
\beqa\label{A.3a}
&&\la 0|  \psi_\r(z)\psi_\s(w)\T(\z) |0\ra=
\frac{\ep_{\r\s}}{(z-\z)^2(w-\z)^2}\ , \\
\label{A.3b}
&&\T(z_1)\T(z_2)=-z_{12}^{-4} + 12  z_{12}^{-5}
\int\limits_{z_2}^{z_1} (z_1-\z)(\z-z_2) \T(\z) \dd\z +
\np{\T(z_1)\T(z_2)}\, \nn
&&   \L(z)=\np{\T(z)^2},\quad
\la 0|  \psi_\r(z)\psi_\s(w)\L(\z) |0\ra=\frac{6}{5}
\frac{\ep_{\r\s}(z-w)^2}{(z-\z)^4(w-\z)^4} \ ;
\eeqa
\end{alphalabel}
\begin{alphalabel}\label{A.4}
\beq\label{A.4a}
\la  0|\psi_\r(z)\psi_\s(w) V_a(\z) |0\ra=
-G_{\r+\s,a}\frac{z-w}{(z-\z)^3(w-\z)^3} \ ,
\eeq
\beqa\label{A.4b}
V_a(z_1)V_b(z_2) &= & G_{ab} \left\{
 z_{12}^{-6} + \frac{18}{z_{12}^{7}}
\int\limits_{z_2}^{z_1} (z_1-\z)(\z-z_2) \T(\z) \dd\z + \cdots \right\} \nn
&&+ 30 \frac{C_{ab}^c}{z_{12}^{8}}
\int\limits_{z_2}^{z_1} (z_1-\z)^2(\z-z_2)^2 V_c(\z) \dd\z +\cdots \ ,
\eeqa
where the non-zero elements of $G_{ab}$ and $C_{ab}^c$ are:
%\addtocounter{labelcounter}{-1}
\beq\label{A.4c}
G_{1 -1}= G_{-1 1}=2,\;\; G_{00}=-1,\;\;\;
C_{1-1}^0=-C_{-1 1}^0=-2,\;\; C_{\pm 1 0}^{\mp 1}=\pm 1.
\eeq
\end{alphalabel}

In fact, the expansion (\ref{A.2}) and the conditions (\ref{A.3})
(\ref{A.4}) are implied by the following result.
%%%%%%%%
\begin{prop} The $4$-point function {\em (\ref{A.1}) } is
reproduced by the OPE,
\beq\label{A.5}
\psi_+(z)\psi_-(w)  =\frac{1}{(z-w)^2} -
\sum_{l=1}^\infty (z-w)^{l-1}  \int\limits_{w}^z P_l(z,w;\z)
\Phi_{l+1}(\z)  \dd\z
\eeq
where the (normalized) weight function,
\beq\label{A.6}
P_l(z,w;\z)=\frac{(2l+1)!}{(l!)^2} \frac{(z-\z)^l (\z-w)^l}{(z-w)^{2l+1}}
\qquad \left( \int\limits_{z_1}^{z_2} P_l(z_1,z_2;\z) \dd\z =1   \right) ,
\eeq
is determined by the condition,
\beq\label{A.7}
\int\limits_{z_1}^{z_2} P_l(z_1,z_2;\z) (\z-z_3)^{-2l-2}\dd\z=
z_{13}^{-l-1} z_{23}^{-l-1} \qquad (z_{ij}=z_i-z_j),
\eeq
while the fields $\Phi_n$ are mutually orthogonal satisfying,
\beqa
\label{A.8}
\la 0| \Phi_n(z_1)  \Phi_m(z_2)|0\ra & = & -C_n (z_{12})^{-2n}\d_{nm}\ ,\\
\label{A.9}
\la 0| \psi_+(z)\psi_-(w)\Phi_n(\z) |0\ra &=&
C_n \frac{(z-w)^{n-2}}{(z-\z)^n(w-\z)^n}, \; n=2,3,\ldots\ ,
\eeqa
and the constants $C_n$ are given by, 
\beq\label{A.10}
C_{n+1}=\frac{n(n+1)}{{2n\choose n}},\quad n=1,2,\ldots
(C_2=1=C_3).
\eeq
Conversely, the fields $\Phi_n$ can be expressed from {\em (\ref{A.5}) }
as composites of $\psi_\pm$:
\beqa\label{A.11}
\Phi_{n+1}(z) &=& -\frac{1}{(2n)!} \lim_{z_1,z_2 \rightarrow z}
\partial_1^n     (- \partial_2)^n  \{ z_{12}^{n+1}
\np{\psi_+(z_1)\psi_-(z_2)} \} \nn
&=& -\frac{(n+1)!}{(2n)!} \sum_{k=0}^{n-1} (-1)^{n-k-1}
{n\choose k}{n\choose k+1}
\np{\partial^k\psi_+(z) \partial^{n-k-1} \psi_-(z)} \ .\quad\quad
\eeqa
\end{prop}
The {\bf proof} is based on the expansion:
\beq\label{A.12}
(1-\eta)^{-2}=\sum_{n=2}^\infty C_n F(n,n;2n;\eta) \ ,
\eeq
for $C_n$ satisfying the recurrence relation,
\[\sum_{l=1}^{n-1} C_{l+1}
\frac{(2l+1)! (n-2)! (n-1)!}{(l!)^2 (n-l-1)! (n+l)!}=1 \ ,
\]
which is solved by (\ref{B.10});
here we use the following
representation for the Hypergeometric function (cf. \cite{bgt,gt}):
\beq\label{A.13}
F(n,n;2n;\eta)=z_{12}^n w_{12}^n \int\limits_{w_2}^{z_2}
\frac{P_{n-1}(z_2,w_2;\z)}{(z_1-\z)^n (w_1-\z)^n}\dd\z\ ,\quad
\eta=\frac{(z_1-w_1)(z_2-w_2)}{z_{12}w_{12}}.
\eeq
It remains to set $\Phi_2=\T$, $\Phi_3=V_0$, $\Phi_4=\h\L$
in order to recover (\ref{A.2}). The OPE of composite fields
(Eqs. (\ref{A.3b}) (\ref{A.4b}) etc.) follow from their
expressions (\ref{A.11}) in terms of the free fields $\psi_\pm$.
Short distance expansions (like (\ref{2.2})) are obtained from here
using:
\beq\label{A.14}
\int\limits_{w}^z P_l(z,w;\z) \Phi_{l+1}(\z)\dd\z= \Phi_{l+1}(w)+
\frac{z-w}{2} \Phi'_{l+1}(w) +O((z-w)^2).
\eeq

%-C----------------------------------------

\sectionnew{Modular Invariants Partition Functions Involving
the Chiral Algebra $\A(\G_{8,4})$}

Let the lattice base vectors $\e^i$ and their dual $\e^*_i$, $i=1,2$
satisfy:
\beq\label{C.1}
G_{2,1}=\left((\e^i|\e^j) \right)=
\left(\matrix{
2 & 0 \cr
0 & 1 \cr}\right) ,\quad \e^1=2\e^*_1, \; \e^2=\e^*_2.
\eeq
The  corresponding $\U1$ currents (which define, for $r=2$, the stress
energy tensor (\ref{B.12}) ) are:
\[J^1(z) \; (\equiv J^{\e^1}(z) )=2J_1(z),\;
\mathrm{and} \;J^2(z)=J_2(z). \]
We are going to classify the ($c=2$) RCFT whose chiral algebra includes
the tensor product  algebra $\A(\G_{8,4})$ where the lattice  $\G_{8,4}$
is spanned by $2\e^1, \ 2\e^2$. We shall continue, however, to identify
the electric charge $\Q$ with (\ref{3.11}), - that is,
\beq\label{C.2}
\Q=\e^*_1  \ , \qquad  \left(|\Q|^2=\h \right).
\eeq
\begin{prop}
There are $5$ $S$ and $T^2$  invariant partition functions for
the $c=1$ chiral algebra $\A(\G_8)$, $3$ such invariants for
$\A(\G_4)$ and hence $15$ factorizable partition functions for
$\A(\G_{8,4})$. The rank $2$ chiral algebra $\A(\G_{8,4})$    also
admits $4$ non-factorizable  modular invariants.
\end{prop}
{\bf Proof.} We first   note that the choice (\ref{C.2})  of the
electric charge requires substituting $\z$ by $2\z$ in the second
argument of the characters $K_\l(.,.;8)$ (\ref{kfuna}).

According to ref. \cite{gan} there are 4 $S$ invariants made out
of $K_\l(\tau,2\z;8)$ (and their conjugate) for which the chiral algebra
is unextended and the number of superselection sectors is 8. They can be
labeled by an integer $l\mod 8$ satisfying $l^2=1 \mod 8$ and are
given by
\beq\label{C.3}
Z_{l,8}(\tau,\z)=\sum_{\l=-3}^4 K_\l(\tau,2\z;8) \overline{K_{l\l}(\tau,2\z;8)}
\quad \mathrm{for} \ \ l=\pm 1, \pm 3.
\eeq
Two of them, $Z_{1,8}$ and   $Z_{-1,8}$  are SL(2,$\Z$) invariant;
the two others,
\[ Z_{3,8}=K_0 \overline{K_0}+  K_4 \overline{K_4} +\left(K_1 \overline{K_3} +
K_{-1} \overline{K_{-3}}+ K_2 \overline{K_{-2}} + c.c .\right) \ ,
\]
and $Z_{-3,8}$ are only $S$ and $T^2$  invariant. There is one more
invariant partition function corresponding to the
$\A(\G_2)\supset\A(\G_8)$ extension of the chiral algebra:
\beqa\label{C.4}
Z_2(\tau,\z) \; (= Z(\G_2\supset\G_8)) &=& 
|K_0(\tau,2\z;8)+  K_4(\tau,2\z;8)|^2\nn 
&&+ \left\vert K_2(\tau,2\z;8)+ K_{-2}(\tau,2\z;8) \right\vert^2.
\eeqa
Due to the identity:
\beq\label{C.5}
K_{2l}(\tau,2\z;4m)+  K_{2l+2m}(\tau,2\z;4m)=K_l(\tau,\z;m) \ ,
\eeq
Eq.(\ref{C.4}) is just the diagonal invariant of the level $1$
su(2) current algebra.

Similarly, there are two modular invariants:
\beq\label{C.6}
Z_{\pm 1,4}(\tau)=\sum_{\l=-1}^2 K_\l(\tau;4) \overline{K_{\pm\l}(\tau;4)}\ ,
\eeq
of  the $\A(\G_4)$  RCFT and one $S$ invariant,
\beq\label{C.7}
Z_1(\tau)=|K_0(\tau;4)+  K_2(\tau;4)|^2 \quad \left(= Z_1(\tau,+2)\right)\ ,
\eeq
corresponding to the  $\A(\G_1)\supset \A(\G_4)$ fermionic extension
of the bosonic chiral algebra $\A(\G_4)$. The products of each of the 3
invariants (\ref{C.6}) (\ref{C.7}) with any of the 5 invariants
(\ref{C.3}) (\ref{C.4}) give the 15 factorizable partition functions.

The 4 non-factorizable $S$ invariants correspond to the
$\A(L)\supset \A(\G_{8,4})$ extension of the original chiral algebra
where $L$ is defined by (\ref{3.6}) and (\ref{3.7}) with $m=2$. They
are related to the partition function (\ref{3.18}) (for $m=2$)
in the same way as the invariants  (\ref{C.3}) are related to the
diagonal one for $\A(\G_8)$:
\beqa\label{C.8}
Z_{l,L}(\tau,\z) &=& \sum_{\l=-3}^4 \left(
K_\l(\tau;4)K_\l(\tau,2\z;8) + K_{\l+2}(\tau;4)K_{\l+4}(\tau,2\z;8)
\right) \times\nn
&&\left(\overline{K_{l\l}}(\tau;4) \overline{K_{l\l}}(\tau,2\z;8) +
\overline{K_{l\l+2}}(\tau;4) \overline{K_{l\l+4}}(\tau,2\z;8) \right)\ ,\nn
&&\qquad l=\pm 1,\pm 3.
\eeqa
The argument of Gannon \cite{gan} then proves that there are no other
$S$ invariants.

{\bf Remark C1.} Note that Gannon\footnote{We
thank T. Gannon for a helpful correspondence
on this point} \cite{gan} (see, in particular,
Example 2 of Section 4) identifies rank 2 charge lattices obtained from
each other by a rotation by $\pi$ and hence views partition functions
differing only in the sign of $l$ as equivalent. In his count there
are, therefore, $3\times 2$ (rather than $5\times 3$) factorizable
and 2 (rather than 4) non-factorizable invariants. We shall see that
some of the partition functions, viewed as equivalent in \cite{gan},
actually differ in their $U-V$ transformation properties.
\begin{prop}
U-invariance leaves us with $9$ factorizable and $2$ non-factorizable
partition functions (among  the $15+4$ ones given by Proposition C1).
If modified by a suitable prefactor of the type appearing in
(\ref{3.13}) they become automatically also $V$-invariant.
\end{prop}
{\bf Proof.} $U$ (i.e., $\z\rightarrow \z+1$) of $Z_{l,8}$ (\ref{C.3})
requires:
\beq\label{C.9}
\l-l\l=0 \ \mod 4 \qquad \mathrm{for} \ -3\leq\l\leq 4 \ ,
\eeq
which is only satisfied for $l=1,-3$. The same is true for the invariants
(\ref{C.8}).

Given that the electric charge vector (\ref{C.2}) has no projection in the
direction of  $\e^2$ one can view a reflection of the $\e^2$    axis
as a symmetry and regard as {\em indistinguishable theories} obtained from
one another by such a reflection (they indeed have identical partition
functions). (Thus we substitute the ``equivalence"
$(\e^1,\e^2) \rightarrow (-\e^1,-\e^2)$ used in \cite{gan} - see Remark C1 -
by   $(\e^1,\e^2) \rightarrow (\e^1,-\e^2)$. )
\begin{prop}
There are $7$ distinct partition functions among the $11$ modular invariant
ones of  Proposition C2. They are given by Theorem 4.1.
\end{prop}
{\bf Proof.} Identifying the two modular invariants (\ref{C.6}) we
remain with $6$ out of $9$ factorizable partition functions. To verify that
$Z_{1,L}(\tau,\z)=Z_{-3,L}(\tau,\z)$ we note that
\beqa
&&K_{-3}(\tau;4)  K_{-3}(\tau,2\z;8)+ K_{-1}(\tau;4)  K_{1}(\tau,2\z;8)=\nn
&&=K_{1}(\tau;4)  K_{5}(\tau,2\z;8)+ K_{-1}(\tau;4)  K_{1}(\tau,2\z;8)\nonumber
\eeqa
differs from
$K_{1}(\tau;4)  K_{1}(\tau,2\z;8)+ K_{-1}(\tau;4)  K_{5}(\tau,2\z;8)$
by just a change of sign in $l$ of the $K_{l}(\tau;4) $ factors.

\def\NP{{\it Nucl. Phys.\ }}
\def\PRL{{\it Phys. Rev. Lett.\ }}
\def\PL{{\it Phys. Lett.\ }}
\def\PR{{\it Phys. Rev.\ }}
\def\CMP{{\it Comm. Math. Phys.\ }}
\def\IJMP{{\it Int. J. Mod. Phys.\ }}
\def\JSP{{\it J. Stat. Phys.\ }}
\def\JP{{\it J. Phys.\ }}

%%%%%%%%%%%%%%%%%%%%%%%%%%%%%%%%%%%%%%%%%%%%%%

\end{document}